\documentclass[aps,pra,groupedaddress,twocolumn]{revtex4}

\usepackage{amsmath}
\usepackage{graphicx}	
\usepackage{bm} 
\usepackage{color}
\usepackage{epstopdf}
\begin{document}
\title{Three-body recombination of two-component cold atomic gases into deep dimers in an optical model}
\author{M. Mikkelsen}
\author{A.~S. Jensen}
\author{D.~V. Fedorov}
\author{N.~T. Zinner}

\affiliation{Department of Physics and Astronomy, Aarhus University, DK-8000 Aarhus C, Denmark}

\date{\today}

\begin{abstract}
We consider three-body recombination into deep dimers in a mass-imbalanced
two-component atomic gas.  We use an optical model where a phenomenological imaginary
potential is added to the lowest adiabatic hyper-spherical potential.
The consequent imaginary part of the energy eigenvalue corresponds to the decay
rate or recombination probability of the three-body system.  The
method is formulated in details and the relevant qualitative features
are discussed as functions of scattering lengths and masses.  We use
zero-range model in analyses of recent recombination data.  The
dominating scattering length is usually related to the non-equal
two-body systems.  We account for temperature smearing which tends to
wipe out the higher-lying Efimov peaks. The range and the strength of the imaginary potential determine positions and
shapes of the Efimov peaks as well as the absolute value of the recombination rate.
The Efimov scaling between
recombination peaks is calculated and shown to depend on both
scattering lengths.  Recombination is
predicted to be largest for heavy-heavy-light systems.  Universal
properties of the optical parameters are indicated.  We compare to
available experiments and find in general very satisfactory agreement.
\end{abstract}

\maketitle

\section{Introduction}

The Efimov effect is very well established and the energies and radii
of the infinitely many three-body bound Efimov states occur in geometric
series corresponding to well defined scaling factors depending on the
masses of the particles.  These states have been searched for and
found, but only indirectly as enhanced decay probabilities when the
two-body interactions are tuned to specific values. Most of these findings 
are for three identical atoms where the geometric scaling factor is $22.7$ 
independent of the atomic mass, see for example the experiments
\cite{Ottenstein2008}-\cite{Huang2014}. The occurrence
conditions for the Efimov effect are that at least two of the three 
two-body subsystems simultaneously interact with infinitely
large scattering lengths, or equivalently with bound states of
zero energy.

The necessary fine-tuning is most easily achieved when at least two of
the particles are identical.  Then obviously at least two subsystems
are identical and therefore simultaneously fine-tuned in an
experiment.  If furthermore one of the particles is distinctly
different, the scaling factors can differ enormously for the resulting
Efimov spectra.  When the system has one light and two heavy
particles, the scaling factor, or the energy spacing, is very
small. This is a huge advantage since many energies then are within the
experimentally accessible range.  The opposite case of one heavy and
two light particles has correspondingly large energy spacing and fewer
accessible states.

The two-component gases, where mass-imbalanced three-body systems are 
likely to recombine in the decay process into a deeply bound dimer and
a third particle, are more difficult to prepare in sufficiently cold 
tunable states.  However, several systems are now investigated experimentally 
\cite{pires2014,tung2014,ulmanis2015} and probably more are in the pipeline.  

It is therefore very timely to investigate these processes
theoretically \cite{zinner2014} in more details and more systematically than done
before, where one of the foci has been on the special case of vanishing
scattering length between the two similar particles, see 
for example \cite{Hammer2010}.  
We shall in this paper only consider negative scattering
lengths where the final state in
the recombination product is a deeply bound dimer, unavailable in usual
zero-range three-body models.  The theoretical formulation is simplest 
for three equal masses, since then the single, and necessarily large, 
scattering length is the only parameter which therefore completely 
determines the properties of the process.  The two-component system is 
much more complicated, since two scattering lengths and one mass ratio
are the necessary parameters to characterize the system.

The scaling factor is very well known as function of masses in the
limit of very large scattering lengths \cite{Fedorov2003,Braaten2006},
that is when both are infinitely large and when only that of the
non-equal particles is large while the third between identical
particles is vanishingly small. These two extreme limits of different
scalings between energies would lead to different periodicities in the
enhanced recombination probability as function of the scattering
lengths.  Any intermediate energy scaling is also possible, and a
smooth transition between these limits arises by varying the third
scattering length between zero and infinity.

One established method is based on a multiple scattering approach
where the final state somehow is simulated although never directly
present.  Recently a novel suggestion appeared to use an optical model
to describe absorption as in nuclear physics and for light waves in
materials.  This formulation can only provide decay, or absorption,
probability without any details of the final states populated in these
processes.  However, this is all we need to describe the measurable
recombination probabilities.  In the present report we shall
elaborate and explore the properties of this optical model.

The overall purpose of the present paper is to discuss three-body
recombination processes in atomic gases.  The emphasis is on
two-component atomic gases where mass-imbalanced recombination
processes are prominent and probably dominating.  This shall be done
by introducing the details of a simple optical model where the real
part is the lowest adiabatic hyper-spherical potential, and the
imaginary part is a square well in hyper-radius.  

The initial zero-range potential must be regularized to provide
physical results. This is achieved by replacing the diverging real
part with a constant at small hyper-radii where the imaginary part
also is finite.  This constant is unimportant for the calculated
recombination rates.  Such an optical model formulation is
phenomenological by nature, but we shall indicate a possible relation
to universally determined values of the model parameters.

The paper is structured with basic definitions in section II, while the
focus on computation of mass-imbalanced three-body recombinations is
formulated in details in section III.  The choice and dependence on
physical as well as model parameters are discussed in section IV.
Precise comparison to available experimental results is the content of
section V. Finally, we summarize a number of conclusions in section
VI, where we also suggest interesting future research projects.

\section{Notation and basic ingredients}

In this section we sketch the details necessary to calculate the
potential which provides the wave functions for recombination
computations.  The first part describes the adiabatic expansion method
and the dependence on masses and scattering length parameters for a
short-range real potential. The second part discusses the extension to
an optical potential by addition of a complex term.

\subsection{Hyper-spherical expansion}

We utilize the formalism developed in \cite{nie01,Fedorov2001} to treat the
3-body problem at low energies.  The particle
coordinates and masses are $\mathbf{r}_i$ and $m_i$, respectively for
$i=1,2,3$.  We describe the three-body system by hyper-spherical
coordinates where the all-important hyper-radius is defined by
\begin{equation}
\label{eq:hypersphericalcoord}
\rho^2=\mathbf{x}_i^2+\mathbf{y}_i^2  = 
\frac{\sum_{i<k}m_i m_k (\mathbf{r}_i - \mathbf{r}_k)^2}
{m(m_1+m_2+m_3)} \; ,
\end{equation}
where $\mathbf{x}_i$ and $\mathbf{y}_i$ are the Jacobi-coordinates, 
\begin{eqnarray}
&& \mathbf{x}_i=\sqrt{\mu_i}(\mathbf{r}_j-\mathbf{r}_k)=\sqrt{\mu_i}\mathbf{r_{x_i}} \nonumber \\ 
&& \mathbf{y}_i=\sqrt{\mu_{jk}}\left(\mathbf{r}_i-\frac{m_j \mathbf{r}_j+m_k\mathbf{r}_k}{m_j+m_k}\right)=\sqrt{\mu_{jk}}\mathbf{r_{y_i}}
\label{eq:jaccord}
\end{eqnarray} 
\begin{equation}
\label{eq:reducedmass}
\mu_i=\frac{1}{m}\frac{m_j m_k}{m_j+m_k} , \;\;\;\;\;\;\; \mu_{jk}=\frac{1}{m}\frac{m_i(m_j+m_k)}{m_i+m_j+m_k}.
\end{equation} 
with an arbitrary normalization mass, $m$, which in the present work
is chosen to be the mass of the lightest particle, unless something
else is explicitly stated.  The first term in the hyper-spherical
expansion is known to give fairly accurate results \cite{Nielsen1998}.
This is provided the excitation energy is less than the energy
difference between states build on first and second adiabatic
potential.  If we extract the radial phase-space factor, the total
wave function, $\Psi$, is approximated by
\begin{equation}
\label{eq:hypersphericalexpansion}
\Psi(\rho,\Omega_{\rho}) \approx \rho^{-\frac{5}{2}} f(\rho)\Phi(\rho,\Omega_{\rho}),
\end{equation}
where $\Omega_{\rho}$ denotes the other five hyper-spherical coordinates.  The
hyper-radial wave-equation to determine $f$ is then given by
\cite{Fedorov2001}
\begin{equation}
\label{eq:hyperradialequation}
\left(-\frac{d^2}{d\rho^2}+\frac{\nu^2(\rho)-\frac{1}{4}}{\rho^2}-\frac{2mE}{\hbar^2}\right)f(\rho)=0 \;,
\end{equation}
where $\nu$ is a function of $\rho$.  The procedure to determine $\nu$
is described in details in \cite{nie01}.  Briefly, the adiabatic
hyper-spherical expansion method amounts to fixing the hyperradius and
solving the three Faddeev equations in the space of the five
dimensionless coordinates, $\Omega_{\rho}$.  All partial angular
momenta are zero corresponding to $s$-wave spherical harmonics
solutions for the angles describing the directions of $\mathbf{x}_i$
and $\mathbf{y}_i$ in all three Jacobi-coordinate sets.  We are
therefore left with the three coupled Faddeev equations describing the
relative sizes of $x_i$ and $y_i$.  For zero-range two-body
interactions these differential equations can be reduced to three
linear algebraic equations with non-trivial solutions only when the
corresponding $3\times 3$ matrix, $M$, has vanishing determinant
\cite{Fedorov2001},
\begin{equation}
\label{eq:detmatrix}
\det M = 0,
\end{equation}
which determines the angular eigenvalue, $\nu$.  The matrix elements
of $M$ are given by \cite{Fedorov2001},
\begin{eqnarray}
\label{eq:diagonal}
M_{ii}&=&\nu\cos\left(\frac{\nu \pi}{2}\right)-\frac{\rho}{\sqrt{\mu_i} a_{jk}}\sin\left(\frac{\nu \pi}{2}\right) \; , \\
\label{eq:offdiagonal}
M_{ij}&=&\frac{2\sin[\nu(\phi_{ij}-\frac{\pi}{2})]}{\sin(2 \phi_{ij})} \;.
\end{eqnarray}
The scattering length, $a_{jk}$, is for the interaction between
particles $j$ and $k$, and the angles, $\phi_{ij}$, are given by
\begin{equation}
\label{eq:massangles}
\phi_{ij}=\arctan\left( \sqrt{\frac{m_k(m_1+m_2+m_3)}{m_i m_j}}  \right) ,
\end{equation}
where $\{i,j,k\}$ is a permutation of $\{1,2,3\}$

\begin{figure}[h]
\centering
\includegraphics[scale=0.15]{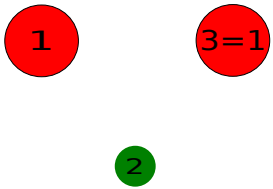}
\caption{Visual illustration of the numbering convention used to
  calculate the adiabatic potentials. Particle $2$ is distinctly
  different from the identical particles $1$ and $3$.}
\label{fig:numberconvention}
\end{figure}

For the simplest case of 3 identical bosons, Eq.(\ref{eq:detmatrix}) 
reduces to $(M_{11}- M_{12})(M_{11} + 2 M_{12}) = 0$, where we only
have to distinguish between diagonal and non-diagonal terms. The second
factor equal to zero is equivalent to the well-known equation for $\nu(\rho)$
\cite{Fedorov2003,Braaten2006}
\begin{eqnarray}
 \nu\cos\left(\frac{\nu \pi}{2}\right)-\frac{8}{\sqrt{3}}\sin(\frac{\nu \pi}{6})
 = \frac{\rho}{\sqrt{\mu} a}\sin\left(\frac{\nu \pi}{2}\right) \;, 
\label{eq:samemassnu}
\end{eqnarray}
where all indices are omitted for this case of identical bosons.  The
other case of interest is two identical and one distinguishable
particle.  We label as illustrated in fig.~\ref{fig:numberconvention},
that is particles 1 and 3 are identical ($m_1=m_3$, $a_{12}=a_{23}$,
$M_{12}= M_{23}$, and $M_{11}= M_{33}$, $M_{13} \neq M_{11}$).  The
characteristic equation then reduces to
\begin{equation}
\label{eq:massimbalancednu}
    (M_{11}- M_{13})(M_{11}M_{22}-2M_{12}^2+M_{22}M_{13})=0 \; ,
\end{equation}
where the last factor provides the general physical solution, which
receives contributions from all three interacting pairs.  Non-interacting
particles $1$ and $3$ correspond to vanishing
scattering length, $a_{13}=0$, where only two interactions contribute.
Then $M_{22} \rightarrow\infty$ and the second factor in 
Eq.(\ref{eq:massimbalancednu}) reduces to $M_{11}+ M_{13}=0$, 
which explicitly amounts to
\begin{equation}
\label{eq:massimbalancednua2zero}
\nu\cos\left(\frac{\nu \pi}{2}\right)+\frac{2\sin(\nu[\phi_{13}-\frac{\pi}{2}])}{\sin(2 \phi_{13})}=\frac{\rho}{\sqrt{\mu_1} a_{12}}\sin\left(\frac{\nu \pi}{2}\right). 
\end{equation}
In all the calculations of $\nu(\rho)$ we only use the smallest
non-trivial solution to the Eqs.(\ref{eq:samemassnu}),
(\ref{eq:massimbalancednu}) and (\ref{eq:massimbalancednua2zero}).
This selects the lowest adiabatic potential in the hyper-spherical
expansion.  The results obtained for $\nu$ depends on the ratios of
hyper-radius to the different scattering lengths.

When $a_i \rightarrow \infty$ the purely imaginary solutions,
$\nu=\nu_0$, become independent of $\rho$, where $\nu_0$ depends on
the masses and the number of contributing interactions.  The radial
equation in Eq.(\ref{eq:hyperradialequation}) has then infinitely many
bound solutions, the Efimov three-body states, with energies, $E_n$,
related by
\begin{equation} \label{energyscale}
\frac{E_{n+1}}{E_{n}}=e^{-2\pi/|\nu_0|}  \equiv \frac{1}{s^2},
\end{equation}  
where we generally refer to $s$ as the Efimov scaling.  We emphasize
that $s$ in particular depends on the number of contributing
subsystems.  In practical comparison with measurements this means that
we can expect $s$ to take any value between the values of the two extreme limits of
two or three contributing subsystems.

\subsection{Optical potential}

We are interested in modelling the recombination for negative values of
all the scattering lengths, that is in the regime where no two-body
bound states exist within the present zero-range model.
We therefore need to introduce the final dimer states populated in the
final state of the three-body recombination process.  However, the only
information we need is the rate of population, or equivalently the
rate at which the three-body system disappears.  No details of the
final states are required, and the optical model is perfect for this
purpose \cite{sie87}.  This model has recently been employed to
describe three-body recombination of identical bosons
\cite{Peder2013}.

We add an imaginary part to the adiabatic hyper-radial potential in
Eq.(\ref{eq:hyperradialequation}) leading to non-hermiticity and
non-conservation of probability.  Thus we can describe the
time-dependent probability reduction as an absorption process, since
recombination is removal of probability from the initial three-body
system.  This is then directly aiming for calculation of the
absorption rate.  The recombination process occurs when all three
particles simultaneously are close in space and two can merge into a
dimer, while the third is necessary to conserve energy and momentum.
The hyper-radius is an appropriate and convenient measure of average
distance between the three particles, see
Eq.(\ref{eq:hypersphericalcoord}).  We therefore modify the potential
in Eq.(\ref{eq:hyperradialequation}) at small distances, $\rho<
\rho_{cut}$, by
\begin{equation}
\label{eq:opticalpotential}
  \frac{2m}{\hbar^2}V(\rho)= \begin{cases}
\frac{\nu^2(\rho)-\frac{1}{4}}{\rho^2}& \text{if $\rho > \rho_{cut}$}\\
\frac{\nu^2(\rho_{cut})-\frac{1}{4}}{\rho_{cut}^2}-V_{imag}\cdot i & \text{if $\rho \leq \rho_{cut}$}
\end{cases} ,
\end{equation}   
where $\rho_{cut}$ and $V_{imag}$ are constants characterizing the
optical potential, $V(\rho)$.  This structure regularizes the
otherwise diverging, for $\rho \rightarrow 0$, real part of the
potential by using a $\rho$-independent constant for $\rho<
\rho_{cut}$.  

The solution, $f(\rho)$, to Eq.(\ref{eq:hyperradialequation}) is
that of a complex square well for $\rho \leq \rho_{cut}$, that is
\begin{equation}
f(\rho)=A \sin(\kappa \rho) \;\;\;\; , \;\;\;\; \kappa=\sqrt{2m(E-V(\rho_{cut}))/\hbar^2},
\label{eq:analyticalsquarewell}
\end{equation} 
where both $\kappa$ and $V$ are complex quantities.  This solution can
conveniently be used as initial condition in the numerical solution
for an integration starting in $\rho_{cut}$.

\begin{figure}[h]
\centering
\includegraphics[scale=0.25]{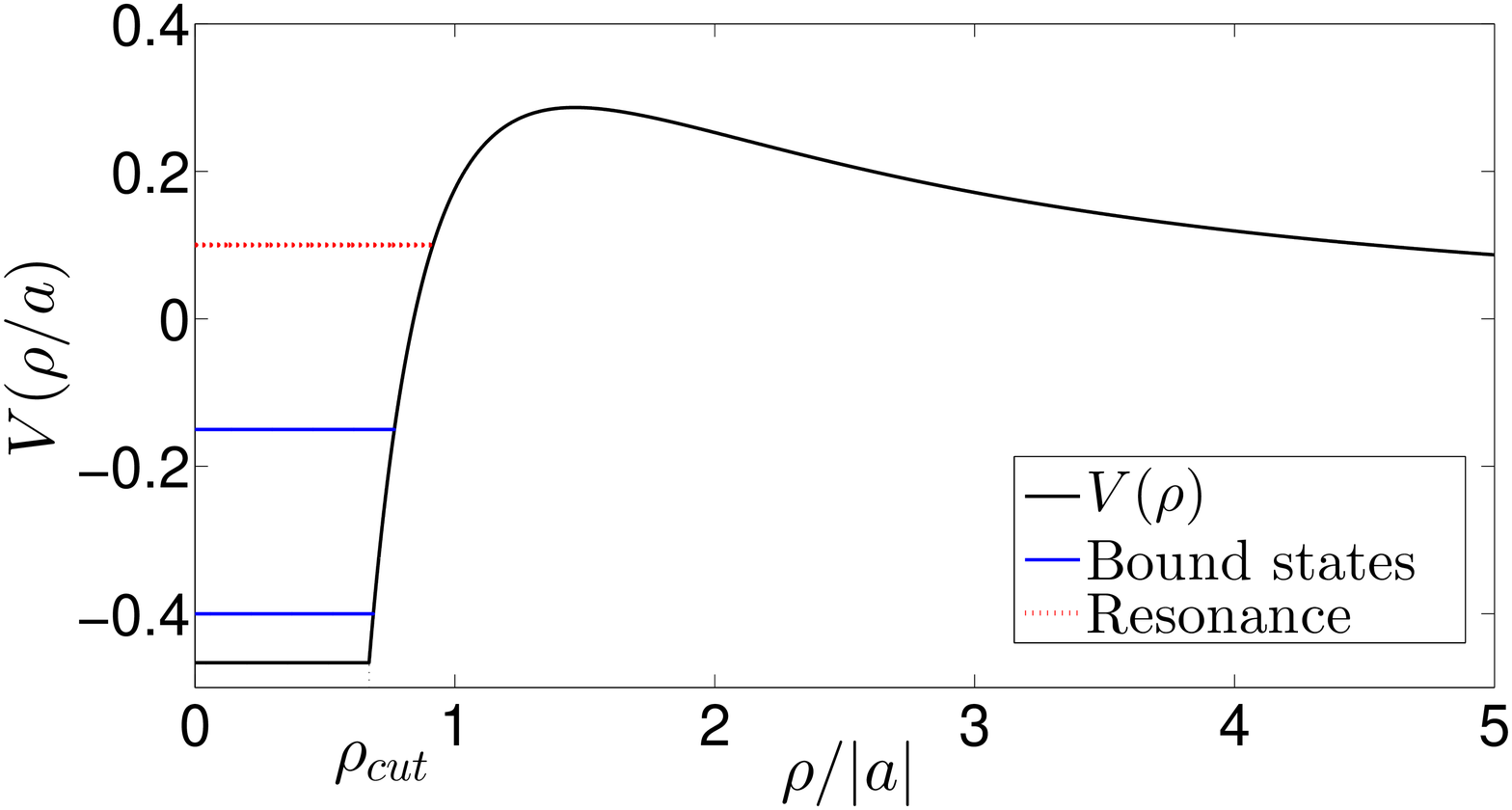}
\caption{A schematic illustration of the real part of the optical
  potential for three identical bosons. Both axes use natural length
  and energy units of $a$ and $\hbar^2/(2ma^2)$.  For values 
  $\rho>\rho_{cut}$ it corresponds to the adiabatic potential for 
  identical bosons.  At smaller $\rho$ it has a constant real value 
  corresponding to the value of the adiabatic potential in $\rho_{cut}$. 
  The blue lines correspond to some of the bound states of the 
  $-1/\rho^{2}$ potential. The dotted red line illustrates a possible 
  resonant state at positive energy. The scale of this illustration is 
  unphysical, since the value of $\rho_{cut }$ in actual calculations 
  tends to be much smaller relative to the value of $\rho$ where the 
  potential has a maximum. }
\label{fig:opticalmodel}
\end{figure}  

A schematic illustration of the real part of the potential is shown
for three identical bosons on fig.~\ref{fig:opticalmodel}.  This
potential has a short-distance attractive region, a barrier of height
$0.287 \hbar^2/(2ma^2)$ located at $\rho =  1.46 a$, and a decrease
towards zero as $15/(4\rho^2)$, where the latter can be seen from
Eq.(\ref{eq:hyperradialequation}) with $\nu=2$ obtained from
Eq.(\ref{eq:samemassnu}), since $\sin(\nu\pi/2)=0$ in the limit of
$\rho \rightarrow \infty$.

The small distance attractive behavior outside $\rho_{cut }$, but
inside the barrier, is proportional to $-1/\rho^2$.  Such a potential
produces a number of bound three-body states, which only is limited by
the finite value of $V$ for $\rho < \rho_{cut }$.  The energies of
these states are related through the scaling in
Eq.(\ref{energyscale}).  Physically we can think of the recombination
process as related to the probability of reaching the absorptive small
distances by tunneling through the barrier.  The recombination
probability is then substantially enhanced when one of these
states has an energy equal to the total three-body energy.

The scattering length is decisive for potential and bound state
energies.  In particular, for three identical bosons we define
$a^{(-)}$ as the value for which a bound state appears with zero
energy.  The recombination probability consequently peaks for the
same, very small, energy close to zero when $a=a^{(-)}$.  The
parameter $\rho_{cut}$ turns out to determine the value $a^{(-)}$,
whereas the strength parameter $V_{imag}$ determines the shape and
size of this recombination peak as a function of $a$.

\section{Three-body recombination}

The potential described in the previous section now has to be used to
calculate the rate of probability disappearing in the different
three-body channels. We first define formally the different channels
and rates, then we describe how to compute these rates, first
by the traditional method employed for identical
particles, and in the last subsection we discuss a new method
intuitively related to the optical model.

\subsection{Rate equations}

For mass-imbalanced recombination the rate equations for the
loss-rates are more complicated than for the mass-balanced case.  For
two components with densities $n_1$ and $n_2$, we can have
three-body systems with one distinguishable and two identical 
particles.  In total we then have 4 different possible recombination processes,
and we can in principle find the time derivative of either of the two
particle densities which leads to 6 different recombination
coefficients, denoted $\alpha$, that is 
\begin{eqnarray}
\dot{n_1}=-\alpha_{111}^{(1)} n_1^3-\alpha_{112}^{(1)} n_2 n_1^2 -\alpha_{221}^{(1)} n_2^2 n_1 ,
\label{eq:rateequationalpha1}
\end{eqnarray}
\begin{eqnarray}
\dot{n_2}=-\alpha_{222}^{(2)} n_2^3-\alpha_{112}^{(2)} n_2 n_1^2 -\alpha_{221}^{(2)} n_2^2 n_1 .
\label{eq:rateequationalpha2}
\end{eqnarray}
Here the indices 1 and 2 refer to the two different kinds of particles
in a two-component gas.  The first term in
Eqs.(\ref{eq:rateequationalpha1}) and (\ref{eq:rateequationalpha2})
corresponds to the recombination coefficient for identical particles,
whereas the last two terms correspond to the two mass-imbalanced
recombinations.

We shall in the following formal derivations often imagine the example
of a Cs-Li gas where Cs is particle 1 and Li particle 2.  Then
$\alpha_{112}^{(1)}$ corresponds to the recombination process between
two particles of type 1 and one particle of type 2, and the rate
relates to the change in the particle density of type 1.  Analogously
for other sub- and superscripts.

The rate equations only describe how many particles disappear from a
three-body system. The final state is not specified.  We could attempt
to further split the mass-imbalanced recombinations into two types
depending on the final structure of which particles form the dimer,
that is
\begin{equation*}
1+1+2\rightarrow1+(12)  \;\;\;\;\;\;\; \text{or}  \;\;\;\;\;\;\; 1+1+2\rightarrow(11)+2.
\end{equation*} 
However, the optical model does not allow distinction between final
states, and there is also currently no way to distinguish the two
channels experimentally.  Thus, $\alpha_{ijk}^{(i)}$ is the full
recombination coefficient corresponding to the sum of these different dimer
productions.

It is obviously much more difficult to find $n_1(t)$ and $n_2(t)$ from
Eqs.(\ref{eq:rateequationalpha1}) and (\ref{eq:rateequationalpha2})
than solving one much simpler equation corresponding to a gas of
identical bosons.  However, we assume the coefficients
$\alpha_{ijk}^{(i)}$ are density and time-independent, and we are able
to calculate these coefficients directly from the radial equation in
the optical model-modified Eq.(\ref{eq:hyperradialequation}).  Also
experimental analyses are more direct by measuring the loss of
individual particles as function of time.

We therefore do not attempt to find the full time dependence from
Eqs.(\ref{eq:rateequationalpha1}) and (\ref{eq:rateequationalpha2}).
It is, however, reassuring to estimate the time-scale at which the
numerically computed values of $\alpha_{ijk}^{(i)}$ make sense.  Let
us first assume that we have a single-species gas corresponding to the
equation $\dot{n_1}=-\alpha_{111}^{(1)} n_1^3$.  The solution is
\begin{equation}
\frac{n_1(t)}{n_1(t_0)}=\frac{1}{\sqrt{1+2\alpha_{111}^{(1)} n^2_1(t_0)(t-t_0)}}.
\label{eq:ratesolutionidentical}
\end{equation}
This square root dependence holds for the single-species gas.  With
typical experimental parameters for the density \cite{Huang2014}, we
find that $n_1$ is reduced by a factor of $2$ over a time period
varying from about half a second to a few nanoseconds as the
scattering length changes from -100$a_0$ to -20000$a_0$, where $a_0$
is the Bohr radius.  We used here the values of $\alpha_{111}^{(1)}$
obtained in the calculations reported in details in a later section of
this paper.

Let us then consider Eq.(\ref{eq:rateequationalpha1}) but
with $n_2$ varying slowly enough to assume it is constant.
If now the $\alpha_{112}^{(1)}$ recombination is dominant, we get the
solution
\begin{equation}
\frac{n_1(t)}{n_1(t_0)}=\frac{1}{1+\alpha_{112}^{(1)} n_1(t_0) n_2(t_0)(t-t_0)}.
\label{eq:ratesolutionapprox1}
\end{equation} 
Under these assumptions, we get a half-life ranging from a few
milliseconds to a few nanoseconds as the $a_{12}$ scattering length
grows from -100$a_0$ to -10000$a_0$.  These numbers correspond to a Cs-Li
gas, where the densities are obtained from the experiment
\cite{pires2014} and the values of $\alpha_{112}^{(1)}$ are from 
calculations reported later in this paper.

If the last term is dominating in Eq.(\ref{eq:rateequationalpha1}) we
get analogously an exponential time dependence, that is 
\begin{equation}
\frac{n_1(t)}{n_1(t_0)}= \exp(-\alpha_{221}^{(1)} n^2_2(t_0)(t-t_0)) .
\label{eq:ratesolutionapprox2}
\end{equation} 
Using the densities obtained from the experiment
\cite{pires2014}, this half-life ranges from about 150 seconds to a few 
milliseconds as the $a_{12}$ scattering length increase from about -100$a_0$ to 
-10000$a_0$. The relatively large half-life here reflects the small values 
of $\alpha_{221}^{(1)}$, which are from calculations reported later in 
this paper. 

We shall return with more discussion on calculated values and relative
sizes of the recombination coefficients in
Eqs.(\ref{eq:rateequationalpha1}) and (\ref{eq:rateequationalpha2}).
In general, our results for the rate coefficients,
$\alpha_{ijk}^{(i)}$, combined with our simple approximate solutions
give realistic time-scales compared to experimental conditions. When analysing
the data in a given experiment, a variety of methods are used. Some involve a 
numerical solution of an equation similar to Eq.(\ref{eq:rateequationalpha1}) \cite{pires2014},
while some are more elaborate taking the experimental variation of the temperature with time 
into account \cite{Roy2013}.

In theoretical calculations we generally find the probability loss 
per unit time for a single three-body system, denoted $J_{ijk}$, which
will be refereed to as the recombination rate. From this the probability
loss of single particles involved in the recombination can be found. 
Let us look at the particle loss, corresponding to the terms in 
Eq.(\ref{eq:rateequationalpha1}). The number of particles $1$ lost per 
unit time, $\dot{N_1}$, from one three-body system is respectively $3J$, $2J$ and $1J$
for the (1-1-1), (1-1-2) and (2-2-1) systems. The total number of 
three-body systems is respectively $\frac{N_1^3}{3!}$, 
$\frac{N_{1}^2 N_2}{2}$ and $\frac{N_1 N_{2}^2}{2}$ for the (1-1-1), 
(1-1-2) and (2-2-1) systems. Equivalent arguments hold for 
Eq.(\ref{eq:rateequationalpha2}) and thus we find the total particle loss
per unit time to be
\begin{eqnarray}
\dot{N_1} &=& -\frac{3}{3!}J_{111} N_{1}^3-\frac{2}{2} J_{112} N_{1}^2 N_2-\frac{1}{2} J_{221} N_1 N_{2}^2 
\label{eq:rateequationJ1} \\
\dot{N_2}&=& -\frac{3}{3!}J_{222} N_{2}^3-\frac{2}{2} J_{221} N_{2}^2 N_1-\frac{1}{2} J_{112} N_2 N_{1}^2 .
\label{eq:rateequationJ2}
\end{eqnarray}
The values $J_{ijk}$ depend on the volume in which the three-body 
systems are confined, and to obtain the $\alpha_{ijk}^{(i)}$ coefficients,
we need to transform to densities using the identity $n_{i}=\frac{N_{i}}{V_{i}}$.
By insertion of this in Eqs.(\ref{eq:rateequationJ1},\ref{eq:rateequationJ2})
and comparison with Eqs.(\ref{eq:rateequationalpha1},\ref{eq:rateequationalpha2})
this  yields the following relation between recombination rates and 
recombination coefficients: 
\begin{align}
\alpha_{111}^{(1)}=\frac{
1}{2}J_{111}V_{111}^ 2 \;\;\;\;\;\;\; , \;\;\;\;\;\;\; \alpha_{222}^{(2)}=\frac{1}{2}J_{222}V_{222}^2 \label{eq:alpha111}   \\
\alpha_{112}^{(1)}=J_{112} V_{112}^2  \;\;\;\;\;\;\; , \;\;\;\;\;\;\; 
\alpha_{112}^{(2)}= \frac{1}{2}J_{112} V_{112}^2  \label{eq:alpha112} \\
\alpha_{221}^{(1)}= \frac{1}{2}\ J_{221} V_{221}^2  \;\;\;\;\;\;\; , \;\;\;\;\;\;\; \alpha_{221}^{(2)}=J_{221} V_{221}^2  \label{eq:alpha221} \;.
\end{align}
The indices are necessary to distinguish between the four different
three-body systems. The quantity $V_{ijk}^2 J_{ijk}$, 
is then independent of the volume, under the assumption that the volume
is sufficiently large. The volume used in theoretical calculations 
and the volume corresponding to specific experiments are not the same, but 
$V_{ijk}^2 J_{ijk}$ should be the same and thus it is the quantity
for which comparisons between experiment and theory is meaningful.
The difference between $V_{ijk}^2 J_{ijk}$ and the recombination 
coefficients is the factors related to the number of three-body systems
and particles lost per three-body recombination, all of which have been 
accounted for in Eq.(\ref{eq:alpha111})-Eq.(\ref{eq:alpha221}). For a 
non-BEC gas there is also a multiplicative symmetry-factor corresponding 
to the number of particle permutations $P_{ijk}$, as described in \cite{Greene2009}, but for a BEC-gas 
this symmetry factor disappears \cite{Kagan1985,Greene2009}. For three 
identical particles $P_{iii}=3!$, while for two identical particles and one 
distinguishable particle $P_{iij}=2!$.    

In the following two sections, two methods of calculating $J_{ijk}$
for any three-body system is derived. In order to ease the notation
the subscripts on $J$ and $V$ are omitted.

\subsection{S-matrix method}

The recombination rate $J$ for a given three-body system is defined
from the missing probability after scattering on the optical potential
in Eq.(\ref{eq:opticalpotential}).  Let us assume the three-body
wave-function asymptotically is expressed in Jacobi coordinates as
a three-dimensional plane wave normalized in a box of volume, $V$.  We
expand this wavefunction in hyper-harmonic free solutions, that is
\cite{Garrido2014}
\begin{eqnarray}
 && \psi = \frac{1}{V} e^{i \mathbf{k_x} \mathbf{x} + i \mathbf{k_y}
    \mathbf{y}} = \nonumber \\ && \frac{1}{V} \frac{(2 \pi)^3}{(\kappa
    \rho)^2} \sum_{\mathcal{K}} i^K \mathcal{Y}^{\ast}_{\mathcal{K}}
  (\Omega_{\kappa})  \mathcal{Y}_{\mathcal{K}} (\Omega_{\rho}) J_{K +
    2} (\kappa \rho), 
    \label{eq:planewavexpansion}
\end{eqnarray} 
where $\mathbf{k_x}$ and $\mathbf{k_y}$ are the Jacobi momenta
characterizing the wave function, $\kappa^2 = (k_x^2+k_y^2)$, and
$\Omega_{\kappa}$ denote the five hyper-angles associated with the
directions of $\mathbf{k_x}$ and $\mathbf{k_y}$. The three-body energy
is then defined by 
\begin{equation}
\label{eq:threebodyenergy}
E = \frac{\hbar^2 \kappa^2}{2m}.
\end{equation}
The arguments of the
two hyperspherical harmonics, $\mathcal{Y}_{\mathcal{K}}$, are related
to the spatial and momentum coordinates, respectively.  The collection
of angular quantum numbers are denoted $\mathcal{K}$, where we only
need to specify the hyper-momentum quantum number, $K$.  The kinetic
energy eigenvalue corresponding to the free wave-function is $K(K+4)$
and at asymptotically large values of $\rho$ related to the eigenvalue
of the adiabatic potentials as $K=\nu(\rho \rightarrow \infty)-2$.
The asymptotic value of $\nu$ for the lowest adiabatic potential is 2,
corresponding to $K=0$.

The radial dependence is given in terms of the spherical
Bessel function, $J_{K+2}$, of order $K+2$.  The hyper-harmonics are
normalized to unity
\begin{equation}
  \int d \Omega \mathcal{Y}^{\ast}_{\mathcal{K}'} \mathcal{Y}_{\mathcal{K}} =
  \delta_{\mathcal{K}' \mathcal{K}} ,
\end{equation}
where the delta-function express that all quantum numbers pairwise
must be equal.  The large-distance asymptotics, $\rho \rightarrow
\infty$, is then obtained from the Bessel function, that is
\cite{Abramowitz}
\begin{equation} \label{asym}
J_{K+2} (\kappa \rho) \xrightarrow[\rho \to \infty]{}  \sqrt{\frac{2}{\pi \kappa \rho}} \frac{e^{i \kappa \rho - i\varphi_K} + e^{- i \kappa \rho + i \varphi_K}}{2},
\end{equation}
where $\varphi_K = (K + 2) \frac{\pi}{2} - \frac{\pi}{4}$.  The two
terms in Eq.(\ref{asym}) correspond to in- and out-going
hyper-sperical waves, respectively.  We now introduce a short-range
optical potential and a unit amplitude on the in-coming wave.  The
absolute value of the out-going amplitude is then asymptotically
allowed to differ from unity.  From Eqs.(\ref{eq:planewavexpansion})
and (\ref{asym}) we then obtain the asymptotic from the $K=0$
wave function, that is
\begin{equation}
\frac{1}{V} \frac{(2 \pi)^3}{(\kappa \rho)^2} \mathcal{Y}^{\ast}_0 (\Omega_{\kappa}) \mathcal{Y}_0 (\Omega_{\rho}) \sqrt{\frac{2}{\pi \kappa \rho}} \frac{S_{11}
  e^{i \kappa \rho - i \varphi_0} + e^{- i \kappa \rho + i \varphi_0}}{2},
  \label{eq:asymptoticK0}
\end{equation}
where $0<|S_{11}|<1$, since the optical potential acts as a sink of
probability. The hyper-radial current is defined by
\begin{equation}
\label{eq:fluxdensity}
  j_{\rho}= - i \frac{\hbar}{2 m} \left(
  \psi^{\ast} \frac{\partial}{\partial \rho} \psi - \frac{\partial}{\partial
  \rho} \psi^{\ast} \psi \right). 
\end{equation}

Using Eqs.(\ref{eq:asymptoticK0}) and (\ref{eq:fluxdensity}) the
missing current, $\Delta j_{\rho} =
j_{\rho}(S_{11}=1)-j_{\rho}(S_{11})$, is calculated for a given value
of $S_{11}$.  This amounts to
\begin{equation}
 \Delta j_{\rho}= \frac{1}{V^2} \frac{\hbar}{m}
  \kappa \left( \frac{2 \pi}{\kappa \rho} \right)^5 (1-| S_{11} |^2)
  |\mathcal{Y}^{\ast}_0 (\Omega_{\kappa}) \mathcal{Y}_0 (\Omega_{\rho}) |^2 .
\end{equation}
To get the probability loss per unit time, $J$, we need to integrate
the missing current over the hyper-surface. The surface 
element must correspond to the physical volume element (see \cite{Garrido2014}). The volume
element is thus given by (see Eqs.(\ref{eq:jaccord}) and 
(\ref{eq:reducedmass})):
\begin{eqnarray}
&& d^3 \mathbf{r_x} d^3 \mathbf{r_y}=\left( \frac{1}{{\mu}_i {\mu}_{j k}} \right)^{3 /
  2} d^3 \mathbf{x} d^3 \mathbf{y}= \nonumber \\
&& m^3 \big(\frac{m_i+m_j+m_k}{m_i m_j m_k}\big)^{3/2}  \rho^5 d
  \rho d \Omega_{\rho} ,
  \label{eq:jacobivolume}
\end{eqnarray}
where we do not have to distinguish between choice of Jacobi coordinates,
since the mass factor is symmetric under exchange of $i$, $j$ and $k$.
To take into account the degeneracy of the initial states for a 
given $\kappa$ we need to average over $\Omega_{\kappa}$, which 
we do by integrating over the angles and dividing by 
$\int d\Omega_{\kappa}=\Omega_5=\pi^3$. For a given $\kappa$ (or energy)
, and under the assumption that we are at large values of $\rho$, we then get
\begin{equation}
J = \int m^3 \big(\frac{m_i+m_j+m_k}{m_i m_j m_k}\big)^{3/2} 
\frac{d \Omega_{\kappa}}{\Omega_{5}} \rho^5 d \Omega_{\rho} \Delta j_{\rho}.
\end{equation}
Exploiting the orthonormality of the hyperspherical harmonics, the missing 
probability per unit time is then
\begin{equation}
J = m^3 \big(\frac{m_i+m_j+m_k}{m_i m_j m_k}\big)^{3/2}\frac{1}{\Omega_{5}} \frac{\hbar \kappa}{{m}} \left( \frac{2\pi}{\kappa} \right)^5 (1-| S_{11} |^2) \frac{1}{V^2}.
\end{equation}
In order to obtain the recombination coefficients we must multiply by $V^2$. This gives
\begin{equation}
 J V^2= m^3 \big(\frac{m_i+m_j+m_k}{m_i m_j m_k}\big)^{3/2}8 \pi^2 \frac{\hbar^5}{m^3} \frac{(1-| S_{11} |^2)}{E^2},
\label{eq:recS}
\end{equation}
which corresponds to the general formula for N-body loss given
in \cite{Greene2009}.

\subsection{Decay-rate method}
In this section we present an alternate way to obtain the recombination 
rate $J$ for the three-body system. This is done by deriving the 
decay rate of bound states in the optical potential.  First we define 
the bound states in a large box with hyper-radius extending from zero to 
$\rho_{max}$. The boundary condition is that the wave function 
is zero at the edge of the box, that is $f(\rho_{max})=0$.  
The eigenvalues for the optical potential are complex numbers
\begin{equation}
E=E_0-\frac{1}{2}i\Gamma.
\end{equation}
The imaginary component of the energy describes the decay rate 
of the probability as seen from the time evolution of the 
wave-function, defined by
\begin{equation*}
|f(\rho,t)|^2 = |f(\rho,t=0)|^2 \exp(-\Gamma t/ \hbar).
\end{equation*}

The decay rates described by $\Gamma=J \hbar$ are three-body energies
determined by a box boundary condition. The overall energy dependence is then
a strong decrease, inversely proportional to the hyper-radial
three-body volume, towards zero as function of box radius
$\rho_{max}$. In order to obtain the recombination coefficient
$\frac{\Gamma}{\hbar} V^2$ we then need to know $V$. 
This volume, $V$, is defined by equating two ways of
calculating the density of three-body states in the hyper-spherical
box extending to $\rho_{max}$.  The first is the formal expression of
integration over given intervals of coordinates and conjugate momenta, 
where $\mathbf{p_x}=\hbar \mathbf{k_x}$ and $\mathbf{p_y}=\hbar \mathbf{k_y}$.
The second is direct numerical calculation of the same quantity from
the solutions to the hyper-radial equation.  The resulting equation is
then
\begin{eqnarray}
 && \int d^3\mathbf{x} d^3\mathbf{y} d^3\mathbf{p_x} d^3\mathbf{p_y} 
 \delta(E-(p_x^2+ p_y^2)/(2m))  = \nonumber \\ \label{eq:phasespace} 
&& V'^2 \Omega_5 \hbar^6 \kappa^5 d\kappa/dE = (2 \pi \hbar)^6 d\nu/dE \;
\end{eqnarray}
where $V'^2=\int d^3\mathbf{x} d^3\mathbf{y}$ is the volume for the
Jacobi coordinates, $\Omega_5 \kappa^5 d\kappa/dE$ is the volume
element per unit energy in momentum space, and $d\nu/dE$ is the
density of states for a given energy, $E$, defined by
Eq.(\ref{eq:threebodyenergy}).  The factor $2\pi \hbar = h $ is
Plancks constant, which is the volume occupied by each quantum state.

Using Eq.(\ref{eq:jacobivolume}) we express $V'$ in terms of the physical
volume, $V$, that is
\begin{equation} \label{volvol}
 V^2 = \int d^3\mathbf{r_x} d^3\mathbf{r_y} = 
 V'^2  m^3 \big(\frac{m_i+m_j+m_k}{m_i m_j m_k}\big)^{3/2} \;.
\end{equation}

Using Eqs.(\ref{eq:phasespace}), (\ref{eq:threebodyenergy}) and
(\ref{volvol}) we then get
\begin{equation} \label{volphase}
V^2= m^3 \big(\frac{m_i+m_j+m_k}{m_i m_j m_k}\big)^{3/2} 
\frac{1}{E^2}\frac{d\nu}{dE}\frac{2}{\Omega_5} \left(\frac{\hbar^2 (2 \pi)^2}{2m}\right)^3 \;,
\end{equation}
where the value of $\frac{d\nu}{dE}$ can be found from solving the
hyper-radial equation numerically.

We emphasize that the lowest hyper-radial equation only accounts for
both total and partial-wave angular momentum zero states, while the
employed phase-space identity includes all states, independent of
angular momentum. The derived relations therefore strongly assume
excitation energies sufficiently small to exclude contributions from
all solutions build on the repulsive higher-lying adiabatic potentials. 

The squared volume $V^2$ scales as $V^2 \propto \rho_{max}^6$, and
meaningful recombination coefficients, independent of box size, are
therefore only achieved when $\Gamma$ is proportional to
$\rho_{max}^{-6}$.  This is  very demanding for numerical
calculations, since convergence only is achieved when $\rho_{max}$ is
larger than the scattering lengths. 

The recombination coefficients, expressed in terms of $\Gamma$,
are then found from
\begin{equation}
J V^2 =m^3 \big(\frac{m_i+m_j+m_k}{m_i m_j m_k}\big)^{3/2}16 \pi^3 \frac{\hbar^5}{m^3} 
\frac{1}{E^2}\frac{d\nu}{dE}\Gamma .
\label{eq:recgamma}
\end{equation}
Eq.(\ref{eq:recgamma}) is equal to Eq.(\ref{eq:recS}) then.
This leads to the immediate conclusion that for an energy $E$ 
corresponding to an allowed eigenvalue the following relation 
between the parameters in the two methods should hold
\begin{equation}
\Gamma=\frac{1}{2\pi}\frac{dE}{d\nu}(1-| S_{11} |^2).
\end{equation}

\subsection{Recombination coefficients and finite temperature}
One immediate conclusion which is readily obtained from Eq.(\ref{eq:alpha111})-
Eq.(\ref{eq:alpha221})is that
\begin{equation} \label{alpgam}
\alpha_{112}^{(2)}= \frac{1}{2} \alpha_{112}^{(1)} \;,\;
\alpha_{221}^{(2)}= 2 \alpha_{221}^{(1)} \;,  
\end{equation}
which is reassuring, as for example one particle of type 2 disappears
for each particle of type 1 in the 1-1-2 recombination. For identical 
particles the mass factor in $J V^2$ (Eq.(\ref{eq:recS}) and Eq.(\ref{eq:recgamma})) 
reduces to $\frac{1}{m_i^3}$, where $m_i$ is the physical mass. Numerically, 
we find that the tunnelling probability $1-|S_{11}|^2$ (for an analytical 
estimate of this, see \cite{pks2013,Sorensen2013}) and the decay rate $\Gamma$ depends on 
the physical mass as $m_i^2$, however, leading to the identity
$\alpha_{111}^{(1)} \approx \frac{m_2}{m_1}\alpha_{222}^{(2)}$. It is 
approximate since the Efimov peaks can have different locations and shapes in the
two systems, leading to non-systematic differences between $\Gamma$ or equivalently $S_{11}$ in
the two systems. This mass-dependence for identical bosons is in correspondence 
with earlier work \cite{Nielsen1999}.  

In order to compare our calculations with experimental data we need to
fold our calculated values of $\alpha_{ijk}^{(i)}(E)$ with a
temperature distribution.  The normalised Boltzmann distribution for 3
particles is given by  \cite{Peder2013}
\begin{equation}
\langle \alpha_{ijk}^{(i)} \rangle_T =\frac{1}{2 (k_B T)^3} \int E^2 e^{-\frac{E}{k_b T}}\alpha_{ijk}^{(i)}(a,E) dE \; ,
\label{eq:temperaturealpha}
\end{equation}
where the $E^2$ factor arises from the three-body phase-space. In order to get
good results it is necessary to calculate $\alpha_{ijk}^{(i)}$ in a
range of energies around $k_b T$, where the integral receives
contributions. 

The highest excitation energies allowed in our low-energy model are
given by the energy difference between states build on the first and
the neglected second and higher adiabatic potentials.  This energy
difference can be estimated by the difference in potential energies at
distances where the states are located.  Since the scattering length
is a measure of the sizes of all these Efimov-like states, we use the
hyper-radius $\rho \approx 10^4 a_0$ to give a lower estimate of the
maximum temperature allowed in realistic calculations.  The energy
difference between first and second adibatic potential is for
interaction free states given by $\hbar^2 K(K+4)/(2m \rho^2)$ where
$K=2$ \cite{nie01}.  The result of these estimates are temperatures of
the order of $\approx 1 \mu K$, which therefore is an upper
temperature limit for realistic calculations.

With the temperature distribution implemented
by Eq.(\ref{eq:temperaturealpha}) the two methods, that is 
the S-matrix and the the decay rate, give 
essentially the same results for the same choice of $\rho_{cut}$ 
and $V_{imag}$, aside from numerical inaccuracies. We shall use 
whichever method is the most convenient in the practical calculations.
For technical reasons it is for example generally more convenient to 
implement the zero-energy limit by the decay rate method, while it is 
easier to implement the temperature distribution for the S-matrix method.

\section{Parameter dependence}

The recombination coefficients for our two-component system depend on three
parameters, that is one mass ratio and two scattering lengths.  We use
the notation from fig. \ref{fig:numberconvention} where the particles
1 and 3 are identical, and 2 is distinctly different.  To illustrate
the effects we investigate first the variation of the adiabatic
potential which is the crucial ingredient in all the calculations.
Before we investigate the dependence of the recombination on physical 
parameters, we investigate the dependence on the optical model 
parameters. Then we investigate the dependence on scattering lengths for 
masses of systems where experimental results are available. Finally we 
compare the different recombinations that can occur in a two-component gas.
The factors corresponding to a non-BEC gas are used for all the recombination 
coefficients.

The numerical calculations are technically simple and employ only
homemade standard programs. First the lowest adiabatic potential is
found numerically from the complex angular eigenvalues, which in tur
is found by solving Eq.(\ref{eq:detmatrix}).  The absorption
probability is then calculated for different energies from the
probability reduction of a plane wave reflected by the optical model
potential.  This involves solving Eq.(\ref{eq:hyperradialequation}),
at a given energy, for the modified potential
Eq.(\ref{eq:opticalpotential}).  This is done by numerical integration
with initial conditions in $\rho_{cut}$ provided by
Eq.(\ref{eq:analyticalsquarewell}).  The result are then fitted to
Eq.(\ref{eq:asymptoticK0}) in order to extract $S_{11}$.  The complex
eigenvalue of the optical potential is calculated by the shooting
method, by requiring that the wave-function is zero at
$\rho_{max}$. The numerical integration is implemented as in the above
$S$-matrix calculations.

\subsection{Adiabatic potentials}

\begin{figure}[h]
\centering
\includegraphics[scale=0.25]{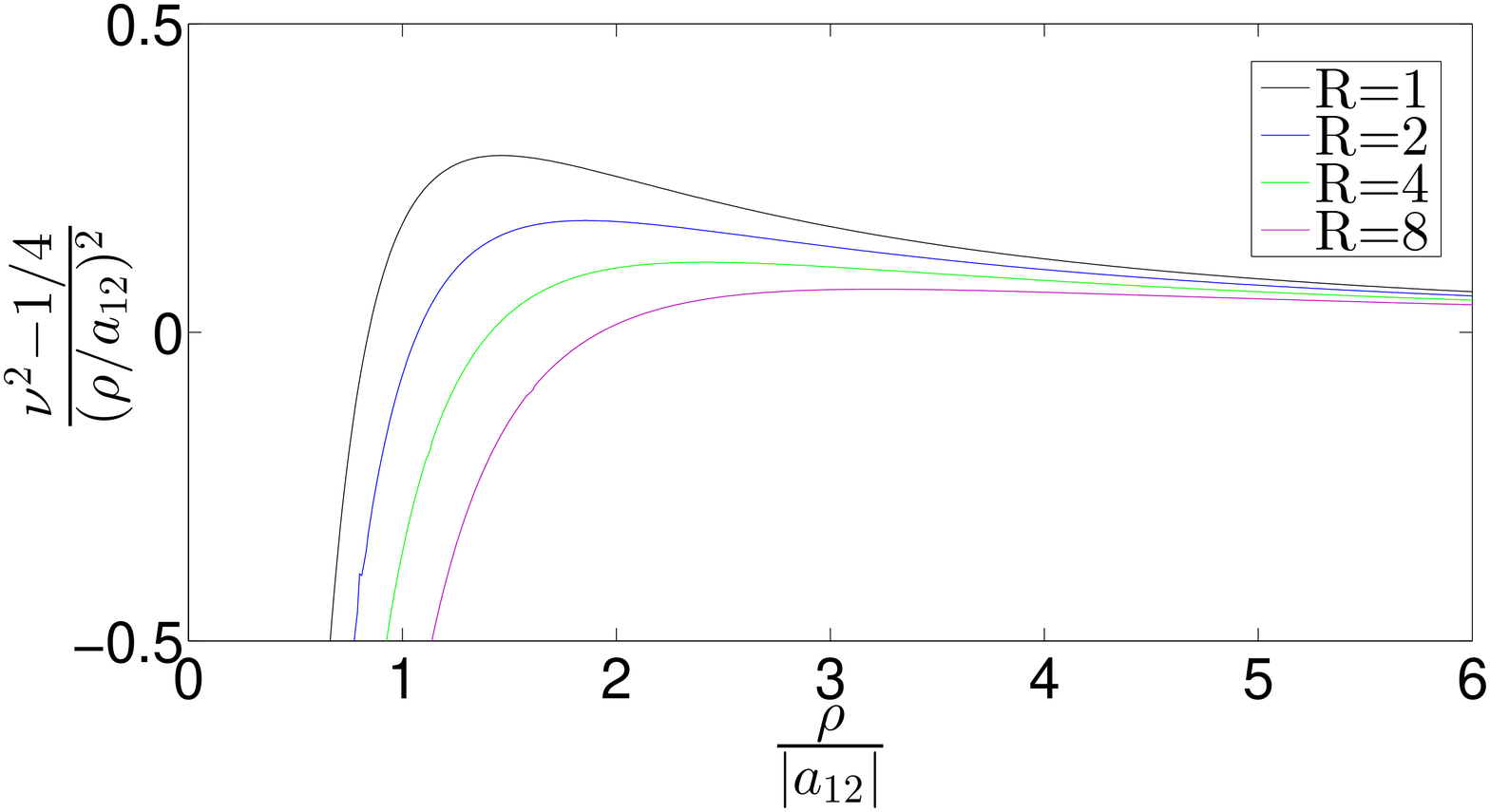}
\includegraphics[scale=0.25]{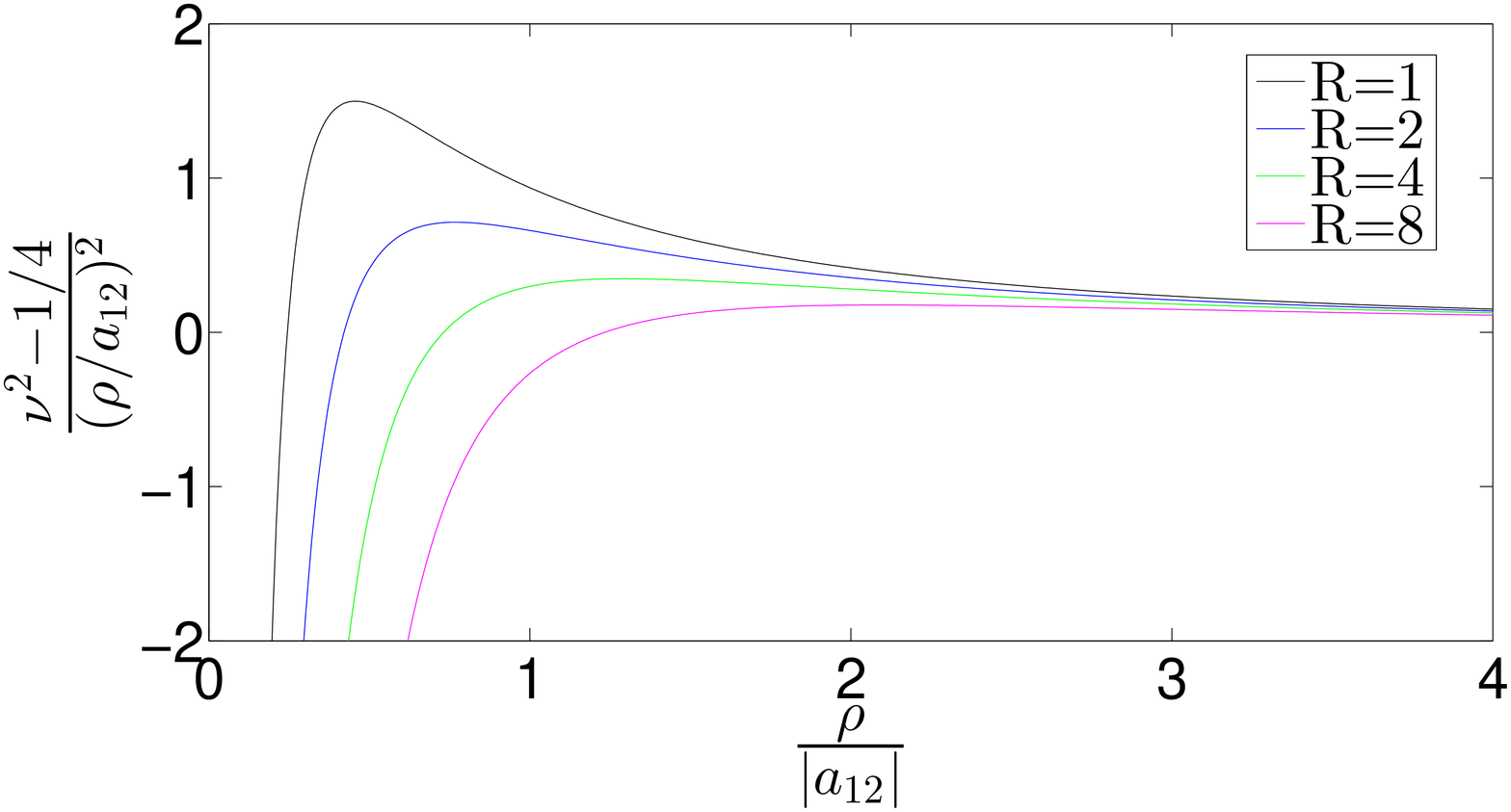}
\caption{The lowest adiabatic potentials as function of
  $\rho/a_{12}$, calculated from Eqs.(\ref{eq:massimbalancednua2zero})
  and (\ref{eq:massimbalancednu}). 
  The lower figure is for $a_{11} = 0$ and the upper figure is for $
  a_{11} = a_{12}$. }
\label{fig:adiabaticpotentialvaryingmassratio}
\end{figure}

The masses only enter as ratios of masses through the $\phi_{ij}$
functions in Eq.(\ref{eq:massangles}), that is for our case this
leaves only one parameter, $R=\frac{m_1}{m_2}$.  We show computed mass
dependence of adiabatic potentials in
fig.~\ref{fig:adiabaticpotentialvaryingmassratio} as functions of
hyper-radius measured in units of the scattering length, $a_{12}$ of the
distinguishable particles.  The two figures show results for a
vanishing and a relatively large scattering length between the
identical particles $a_{11}$.  We see that all potentials
asymptotically approach $\frac{15/4}{\rho^2}$, as in the case of
identical particles.

For both values of $a_{11}$ we find that increasing $R$ leads to lower
barrier height, and a shift to smaller hyper-radii of barrier position
and $\rho$-value, where the potential is zero.  The
dependence is strongest for two heavy and one light particle, that is
for $R>1$. Continuing to $R<1$ would give very little variation in the
potentials compared to  the $R=1$ curve.  Increasing $|a_{11}|$ from zero
reduces the peak heights and move the potentials towards larger
hyper-radii.

\begin{figure}[h]
\centering
\includegraphics[scale=0.25]{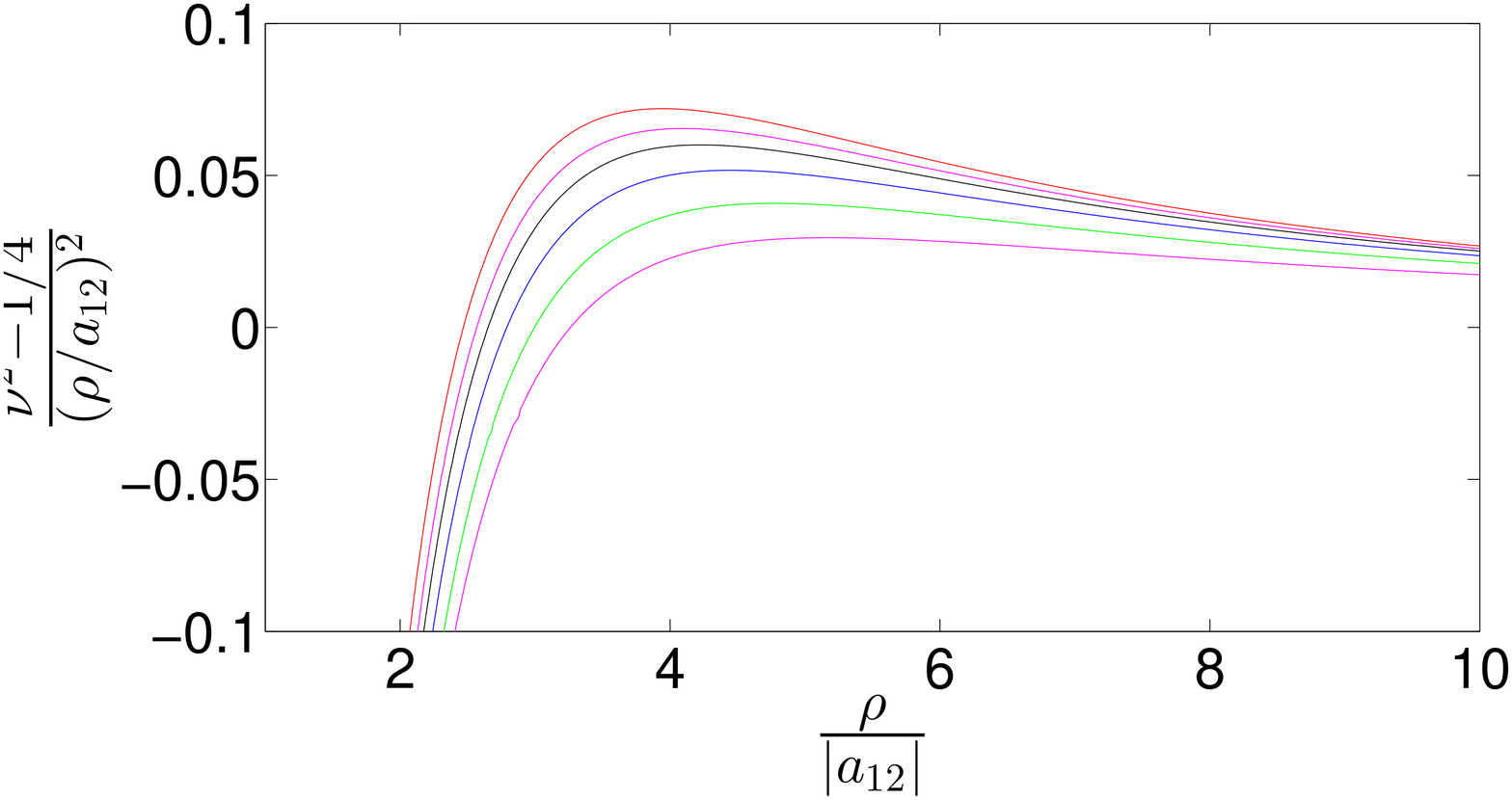}
\caption{The lowest adiabatic potentials as function of
  $\rho/a_{12}$. The red graph corresponds to
  Eq.(\ref{eq:massimbalancednua2zero}). The other graphs,
  in descending order, corresponds to
  $a_{11}=0.08,0.16,0.32,0.64,1.28 a_{12}$ where
  $a_{12}$ is negative, calculated from
  Eq.(\ref{eq:massimbalancednu}).  }
\label{fig:varyinga2adiabatic}
\end{figure}

Experimental data are available for a Cs-Li gas where the mass ratio
is $R = 22.28$. The dependence on the scattering length,
$a_{12}$, is shown in fig.~\ref{fig:varyinga2adiabatic} for different
values of $a_{11}$.  The potential is rather small for this value of
$R$, but still with a distinct barrier where the height decreases with
increasing values of $|a_{11}|$.

\begin{figure}[h]
\centering
\includegraphics[scale=0.25]{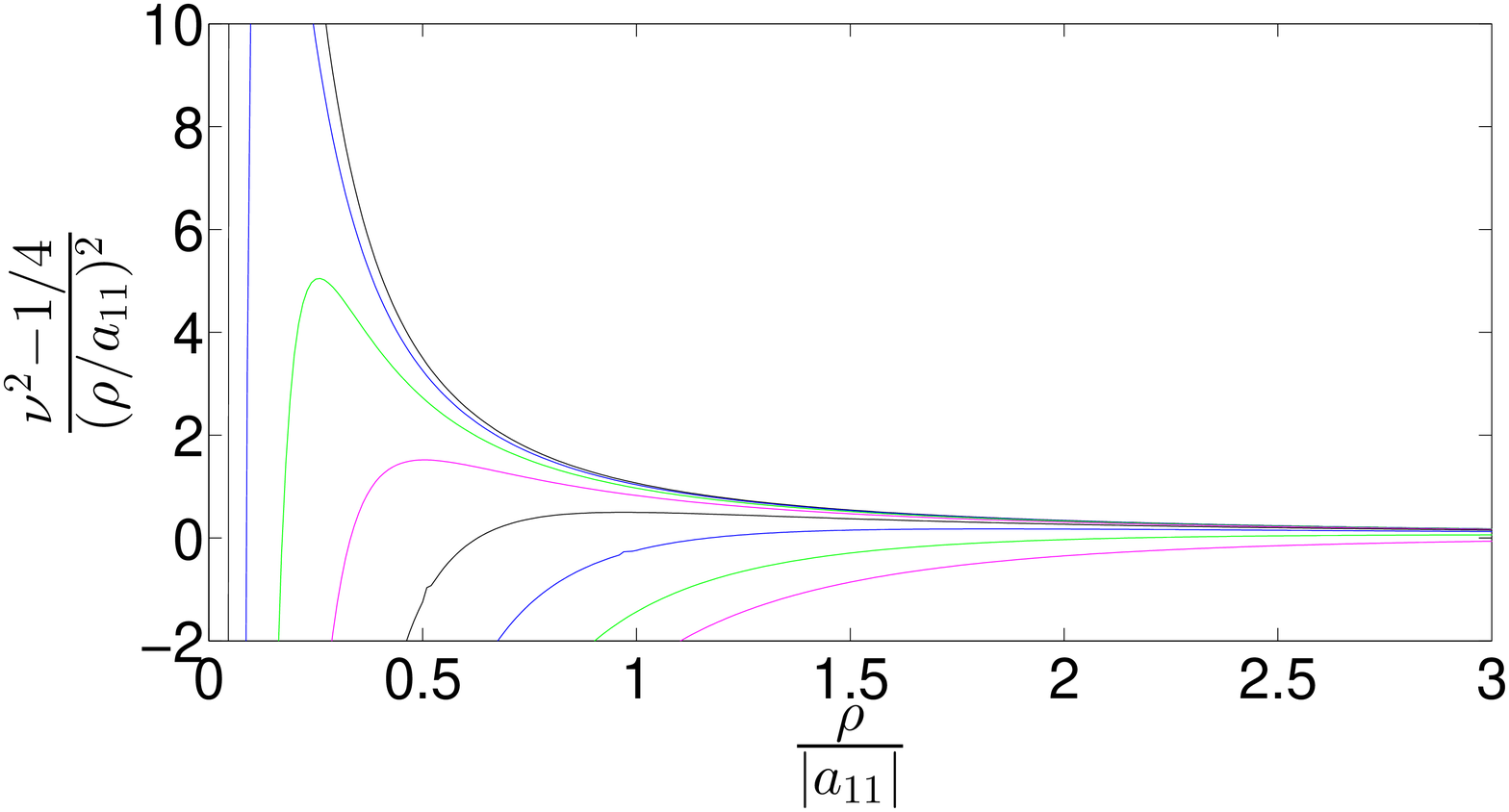}
\caption{The lowest adiabatic potentials as function of
  $\rho/a_{11}$. These graphs, in descending order, corresponds to
  $a_{12}=0.01,0.02,0.04,0.08,0.16,0.32,0.64,1.28 a_{11}$
  where $a_{11}$ is negative, calculated from 
  Eq.(\ref{eq:massimbalancednu}).}
\label{fig:varyinga1adiabatic}
\end{figure}

To complement we show the dependence on $a_{11}$ in
fig.~\ref{fig:varyinga1adiabatic} for different values of $a_{12}$.
The barrier decreases with increasing $a_{12}/a_{11}$ from a rather
large value for vanishing $|a_{12}|$, towards rather small values. For 
large values of $|a_{12}|$ the barrier has become so small and moved
so far to the right, that it is no longer visible on the scale 
chosen for the figure.

These figures show that increasing $|a_{11}|$ and $|a_{12}|$ has the
same qualitative effect as increasing the scattering length in the
case of identical bosons for the optical model, see
fig.~\ref{fig:opticalmodel}. It decreases the height of the barrier
and moves the location of its maximum to the right.

\subsection{Optical model}
\begin{figure}[h]
\centering
\includegraphics[scale=0.25]{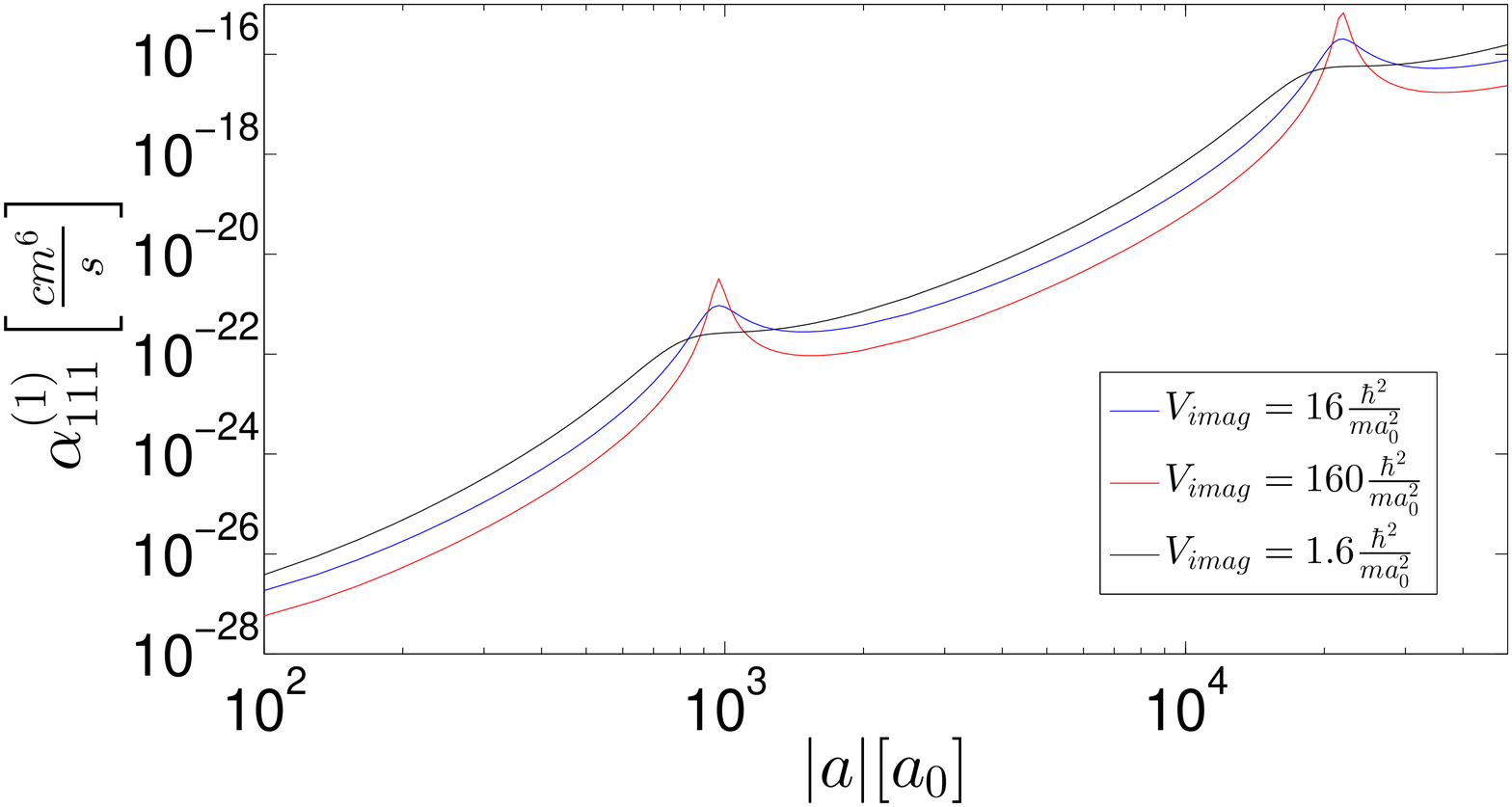}
\caption{The recombination coefficient for Cs-Cs-Cs as function of the
  scattering length $a$ for different choices of $V_{imag}$, at $\rho_{cut}=1.4 a_0$  }
\label{fig:optparam}
\end{figure}

The real potential is completely determined for zero-range two-body
interactions in terms of scattering lengths and masses.  The box-like imaginary
potential has hyper-radius and strength as the two phenomenological
parameters.  This radius can only assume discrete values determined by
the requirement that the scattering length is reproduced for the
occurrence of a specific recombination peak. The corresponding
strength correlatedly varies shape and size of the chosen peak. On 
fig.~\ref{fig:optparam} the results of varying $V_{imag}$, while 
$\rho_{cut}$ is a constant value of $1.4 a_0$ are shown. 

We see that increasing the size of $V_{imag}$ makes the peaks more 
pronounced while also making the absolute value for the rest of the graph
somewhat smaller. Decreasing the size of $V_{imag}$ has the opposite effect.
This is the exact same qualitative effect which was seen when varying
the absorption parameter $n_{*}$ describing deeply bound states in 
\cite{Hammer2010}.

From one set of parameters another equally good choice is found by
scaling the square well hyper-radius up or down by the Efimov factor
and at the same time scaling the strength in opposite direction by the
square of the same Efimov factor. Using these parameters reproduces
exactly the same shape of the Efimov resonance and the same absolute value 
of the recombination. This means that $V_{imag} \rho_{cut}^ 2$ for a given
peak has a constant value where $\rho_{cut}$ 
can discretely vary by the Efimov factor. The largest allowed hyper-radius
must be at a hyper-radius where the real potential precisely has the
inverse radial square behavior. The lowest value is only limited by
the requirement that it must be finite, since otherwise the already
removed zero-range divergency reappears. In practical calculations
we generally pick values in the order of $a_0$ to be certain that 
we do not approach any of these limiting cases.

\subsection{Recombination}

For identical particles we know that the recombination coefficient is
proportional to the fourth power of the scattering length.  For a
two-component system this relation must be extended to account for two
different scattering lengths as well as for a mass-ratio dependence.
We shall illustrate with two mass ratios $R =22.2$ and $R=2.23$,
corresponding to experimentally realizable systems, that is
$^6$Li-$^{133}$Cs \cite{pires2014,tung2014} and $^{39}$K-$^{87}$Rb
gases.  We used these isotopes to have well-defined mass ratios in the
calculations.  The characteristic small mass variations between
isotopes would only marginally change the results, provided the
boson-fermion characters remain the same or become irrelevant as when
only one identical atom participates in the process.  Here we shall
only investigate the pure mass dependence.

The dominating recombination coefficient is related to $\alpha_{112}^{(1)} =
2 \alpha_{112}^{(2)}$, where label 2 corresponds to the light particle.
This means that we here consider the heavy-heavy-light three-body systems
with mass ratios, $R =22.2$ and $R=2.23$.  We calculate all the
recombination coefficients in the limit of zero three-body energy.
 
The periodic structure of enhanced recombination occurs each time the
scattering length is multiplied by the Efimov scaling factor,
$s=\exp(\pi/|\nu(\rho=\infty)|)$, found for infinite scattering lengths
of the contributing systems.  This scaling parameter depends first of
all on the mass ratio.  The Efimov effect requires that $|a_{12}| =
\infty$, while $|a_{11}|$ can assume any finite or infinite values.  The
results for the two limiting cases, $|a_{11}|=0$ and $\infty$, are given in
table \ref{tab:massdependency} as a function of mass ratio, $R$.
For large $R$ the two cases are almost identical, but there is a
big difference for small mass-ratios and this trend continues for
$R<1$.  If all three scattering lengths are infinitely large the scaling
for $R=1$ is $s=22.7$ and for small $R$ approaches a constant of $s=15.7$. For a 
more detailed discussion of the two different cases see \cite{Braaten2006}.

\begin{table}[h]
\scriptsize 
\centering
 \caption{Efimov scaling factor $s$ for different values of R}
\label{tab:massdependency}
    \begin{tabular}{|l|l|l|l|l|l|l|}
    \hline
    $R=\frac{m_1}{m_2}$ & 1       & 2        & 5       & 10     & 15     & 20         \\ \hline
    $a_{12}=a_{11}=\infty$ & 22.7 & 20.8  & 13.7 & 8.50 & 6.30 &  5.14 \\ \hline
    $a_{12}=\infty,a_{11}=0$ & 1986    & 153.8 & 23.3 & 9.76 & 6.64 & 5.25  \\ \hline
    \end{tabular}
\end{table}

The influence of $a_{11}$ can conveniently be studied through the
recombination of the two chosen systems.  For $R =22.2$
the Efimov scaling only varies between 4.8766 for $a_{11}=0$ and
4.7989 for $|a_{11}|=\infty$ and the peak positions should therefore
remain. In contrast the scaling between peaks for $R=2.23$ should move
between the extreme limits of 121.1 for $a_{11}=0$ and
20.28 for $|a_{11}|=\infty$ as a function of $a_{11}$.

\begin{figure}[h]
\centering
\includegraphics[scale=0.25]{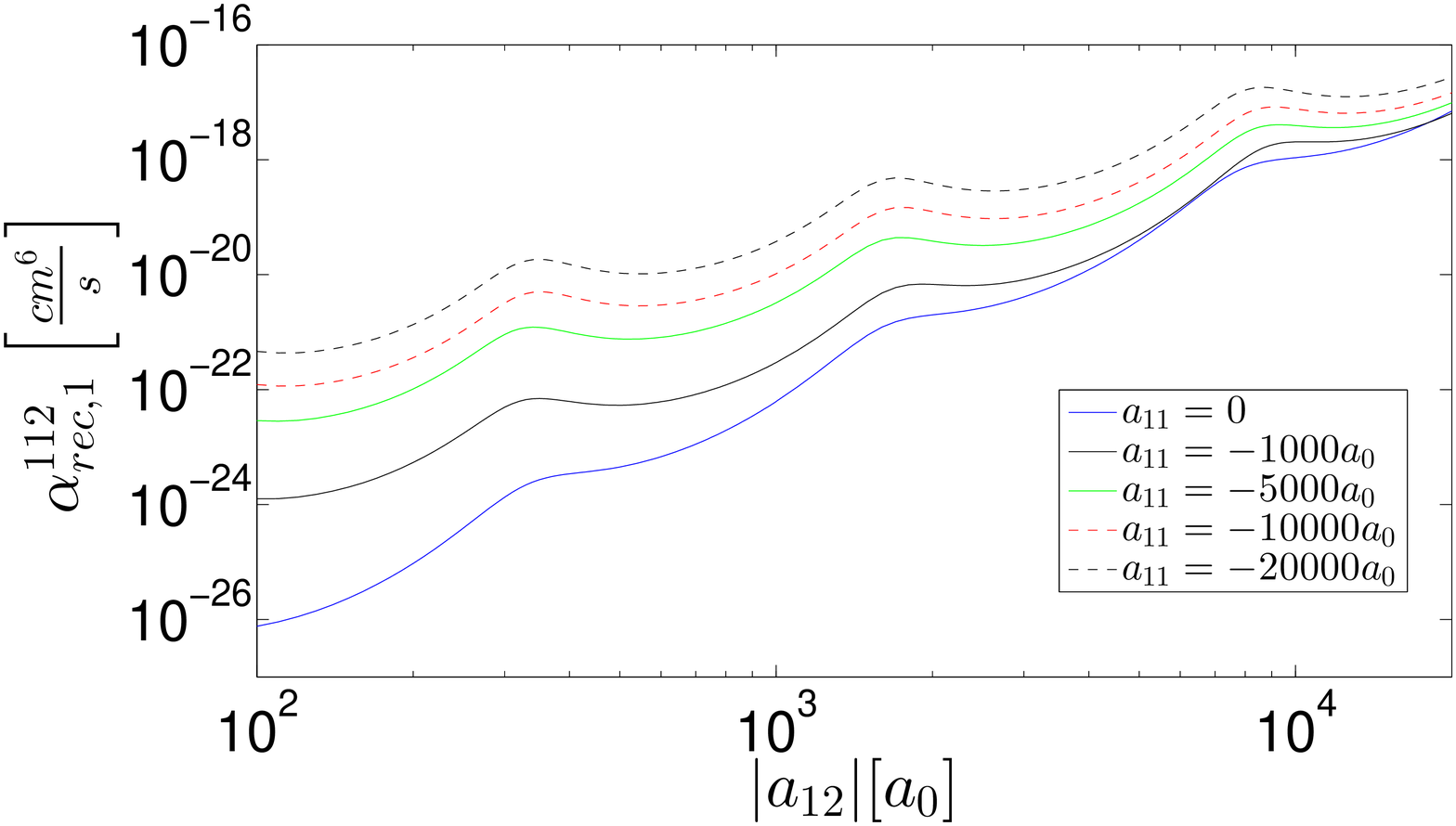}
\caption{The recombination coefficient, $\alpha_{112}^{(1)}$, as a function
  of $a_{12}$, for different values of $a_{11}$, where $R =22.2$.  The parameters
  of the optical potential are $V_{imag}=68 \frac{\hbar^2}{m a_0^2}$,
  with $\rho_{cut}$ in the interval, 0.24-0.32$a_0$ as $a_{11}$ varies 
  from 0 to -20000$a_0$, adjusted to maintain the position of the first 
  Efimov peak at $a_{12} \approx -320a_0$.}
\label{fig:varyinga2forCsLi}
\end{figure}

Let us first focus on the recombination coefficient, $\alpha_{112}^{(1)}$,
for a large mass ratio of $22.28$. We calculate $\alpha_{112}^{(1)}$ as a 
function of $a_{12}$, for different
constant values of $a_{11}$. The strength of the optical potential 
is $V_{imag}=68\frac{\hbar^2}{m a_0^2}$, whereas $\rho_{cut}$ is 
adjusted slightly to reproduce the peak position $|a_{1}^{(-)}|
\approx -320 a_0$, corresponding to the experimental peak position 
of Cs-Cs-Li \cite{pires2014,tung2014}, for the different values of $a_{11}$.

The main results of these calculations are summed up in
fig.~\ref{fig:varyinga2forCsLi}, where we show the coefficient as a
function of $a_{12}$ for different values of $a_{11}$.  The most
striking feature of fig.~\ref{fig:varyinga2forCsLi} is that the
different values of $a_{11}$ lead to a different overall dependence of
$\alpha_{112}^{(1)}$ as function of $a_{12}$. For small values of
$|a_{11}|$ there is a $a_{12}^4$ dependence corresponding to the $a^4$
scaling for identical bosons, but for bigger values of $|a_{11}|$ this
relation is no longer valid.  As $|a_{11}|$ grows bigger, the
recombination is enhanced, which is most visible at smaller values of
$|a_{12}|$.  This is consistent with larger values of each of the
scattering lengths, $|a_{12}|$ and $|a_{12}|$, lowering the potential
barrier, making it easier to reach small distances and thereby
enhancing the recombination.

We also know that $a_{12}$ has a stronger influence on the potential
than $a_{11}$, which corresponds well with $a_{12}$ being the most
important parameter for the recombination shown in
fig.~\ref{fig:varyinga2forCsLi}.  Specifically, we see that
$\alpha_{112}^{(1)}$ changes more as $a_{12}$ runs from 0 to -20000
$a_0$, than it does when $a_{11}$ varies from 0 to -20000~$a_0$.

All in all, for realistic values of the scattering lengths, we can
view $a_{11}$ as moderating the $a_{12}$ dependence and simultaneously
enhancing the recombination.  More violent changes of $a_{11}$ cannot
be excluded when the interactions are controlled by the Feshbach
resonance technique.  However, this is not the case in any of the
physical systems discussed here.

\begin{figure}[h]
\centering
\includegraphics[scale=0.25]{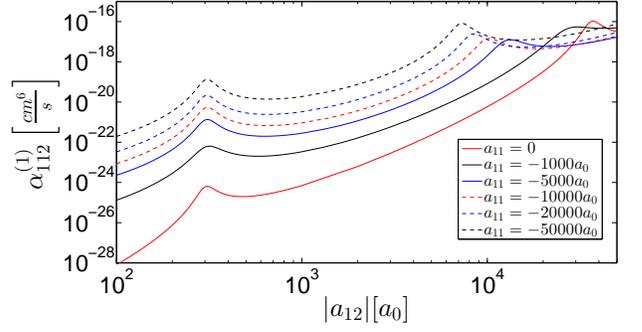}
\caption{The recombination coefficient, $\alpha_{112}^{(1)}$ as a
  function of $a_{12}$ calculated for different values of $a_{11}$,
  where $R =2.23$. The strength of the potential is as in
  fig.~\ref{fig:varyinga2forCsLi} while $\rho_{cut}$ is adjusted to
  give one peak at about $a_{12} \approx -300 a_0$.  }
\label{fig:varyinga2forRbK}
\end{figure}

We now move to the mass ratio $R=2.23$ where the Efimov scaling has a 
strong dependence on $a_{11}$ as seen from the extreme limits given in 
table \ref{tab:massdependency}. The relative positions of the recombination
peaks must then vary with the finite value of $a_{11}$.  Again we calculate 
$\alpha_{112}^{(1)}$ as a function of $a_{12}$ for different values of $a_{11}$.
The results of such a series of calculations are shown in
fig.~\ref{fig:varyinga2forRbK}, where $\rho_{cut}$ is adjusted to
produce the same somewhat arbitrary peak position at about $|a_{1}^{(-)}|
\approx -300 a_0$ for all $a_{11}$.  The next Efimov peak then has to move
in the interval of $a_{12}$ from $20.28 \cdot a^{(-)} \approx - 6000
a_0$ and $121.1 \cdot a^{(-)} \approx - 36000 a_0$.

\begin{table}[h]
 \caption{The ratio, $a_{2}^{(-)}/a_{1}^{(-)}$, between the $a_{12}$
   scattering lengths for the first two peaks as
   function of $a_{11}$ when $R=2.23$. The last column give the related
   $\rho_{cut}$ values. } \centering
    \begin{tabular}{|l|l|l|l|l|l|l|}
    \hline
  $a_{11}$ $[a_0]$                 & 0  & -1000  & -5000 & -10000 & -20000 & -50000 \\ \hline
  $\frac{a_{2}^{(-)}}{a_{1}^{(-)}}$ & 119.4 & 96.9 & 41.9 & 32.9 & 27.7 & 23.9 \\ \hline
  $\rho_{cut}[a_0]$ 		 & 1.21 & 0.97 & 1.14 & 1.18 & 1.2 & 1.22 \\ \hline
  \end{tabular}
  \label{tab:scalingfordifferenta2}
\end{table}

The most noticeable thing in fig.~\ref{fig:varyinga2forRbK} is that
the location of the second Efimov resonance changes with the finite
value of $a_{11}$.  This variation of the ratio between $a_{12}$
values of second and first peak is shown in table
\ref{tab:scalingfordifferenta2} for different $a_{11}$ values.  The
rather modest variation of $\rho_{cut}$ is also shown in this table.
We notice the correct continuous variation between the two extreme
limits of the Efimov scaling with $a_{11}$, although the value of
$\rho_{cut}$ for $a_{11}=0$ reflects the singularity moving between 
Eqs.(\ref{eq:samemassnu}) and (\ref{eq:massimbalancednua2zero}).

We emphasize that small changes of $\rho_{cut}$ is fine-tuning to
maintain the first peak at the same position.  The much larger shift
of the second peak is entirely due to the variation of $a_{11}$, and
essentially completely independent of the parameters of the imaginary
potential.  A substantial change in the second peak position requires
a value of $a_{11} \approx -5000 a_0$.  However, the change is fairly
gradual and there is no "magic value" where $a_{11}$ suddenly begins
to contribute.  The ideal Efimov scaling for three resonant
interactions of 20.28 is essentially reached numerically when
$a_{11}=-50000 a_0$.

The second prominent feature in fig.~\ref{fig:varyinga2forRbK} is that
a finite value of $a_{11}$ modifies the overall dependence of
$\alpha_{112}^{(1)}$ as a function of $a_{12}$.  This is the same
qualitative feature as seen in fig.~\ref{fig:varyinga2forCsLi} for the
large mass ratio $R =22.2$. Finite values of $a_{11}$ leads to a higher 
absolute value of $\alpha_{112}^{(1)}$ within a huge $a_{12}$ interval, 
$a_{12}\in[0,-20000a_0]$.  This means that increasing the size of $|a_{11}|$
enhances the recombination coefficient, like we found for the $R =22.2$ system.

\subsection{Different recombination processes}

We have so far only considered recombination from three-body systems
with two heavy and one light particle such as Rb-Rb-K and Cs-Cs-Li.
However, the same two-component gas can also decay by the other 3
combinations, two light and one heavy particle (Rb-K-K, Cs-Li-Li), and
3 identical heavy (Rb-Rb-Rb, Cs-Cs-Cs) or light particles (K-K-K,
Li-Li-Li).  In order to estimate the importance of the terms in
Eqs.(\ref{eq:rateequationalpha1}) and (\ref{eq:rateequationalpha2}),
it is necessary to calculate all these recombination coefficients.

To make a realistic comparison we first consider the Cs-Li gas, where
the Efimov resonances are fixed by experimental data for Cs-Cs-Li
\cite{pires2014}, Cs-Cs-Cs \cite{Berninger2011} and Li-Li-Li
\cite{Dyke2013}, and the parameters of Li-Li-Cs are assumed to be the
same as for Cs-Cs-Li.  We assume that $a_{11}=a_{22}=0$. This would 
underestimate the recombination coefficients as shown in the previous 
subsection, but it allows on the other hand a clean comparison where 
the fourth power dependence applies, $\alpha_{ijk}^{(1)} \propto a_{12}^4$.  
Actual correlated finite values of these scattering lengths obtained 
through the Feshbach technique could then be important and quantitatively
alter the comparison.

\begin{figure}[h]
\centering
\centering
\includegraphics[scale=0.25]{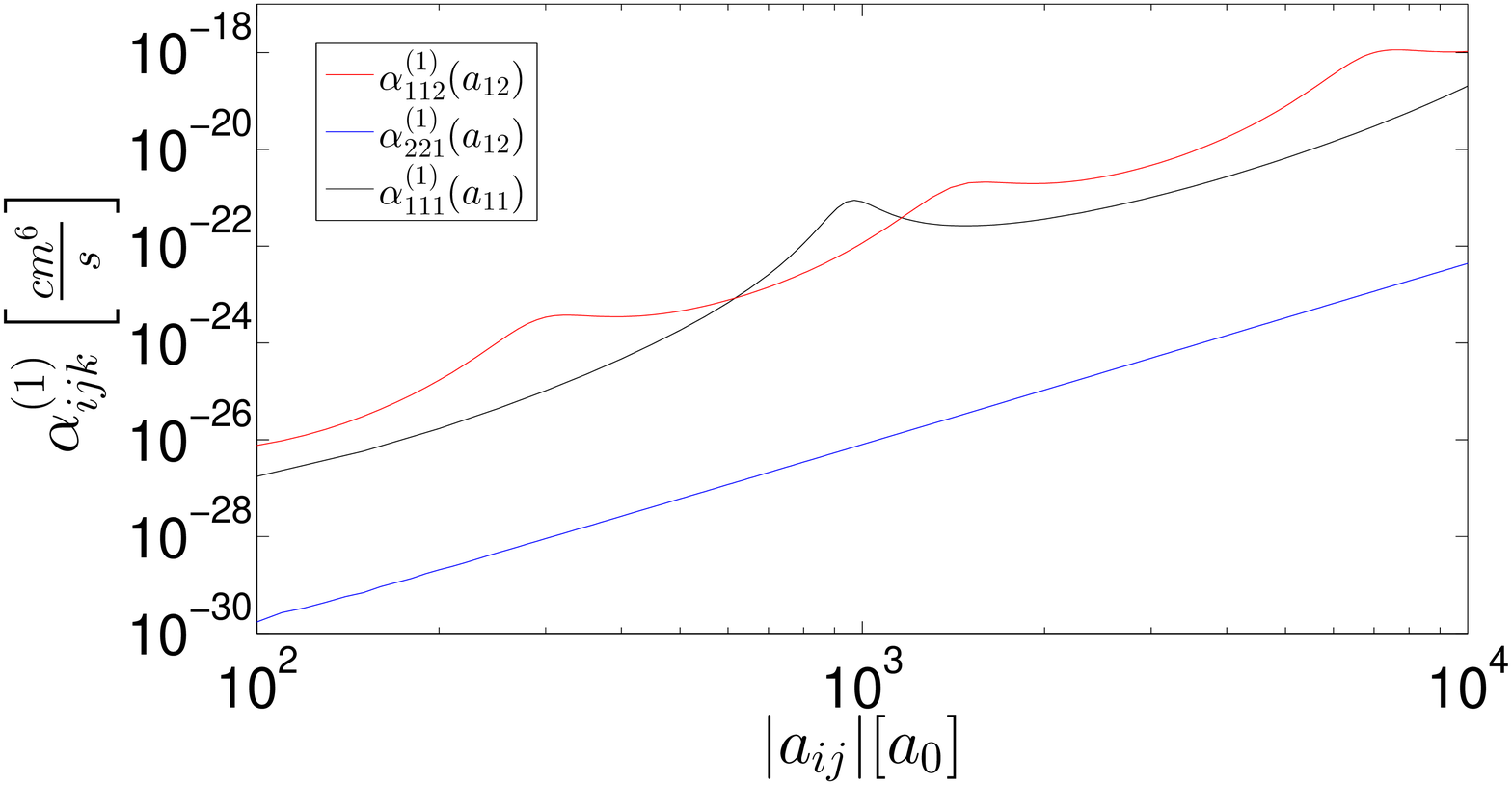}
\includegraphics[scale=0.25]{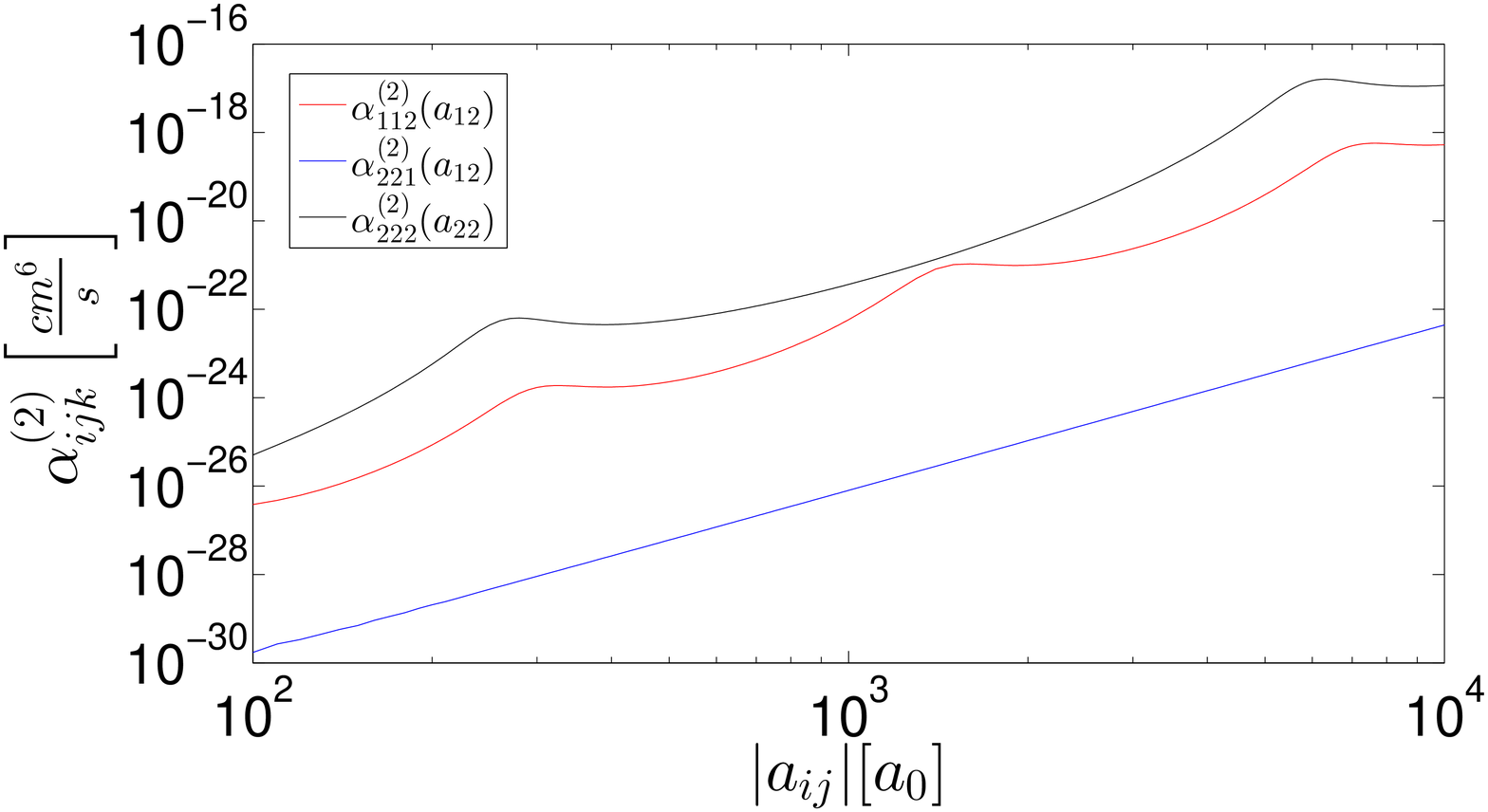}
\caption{The top shows the recombination coefficients resulting in Cs
  loss, and the bottom shows the recombination coefficients resulting
  in Li loss, both as a function of $a_{12}$.  The optical parameters
  in the calculations are $\rho_{cut}=0.20 a_0$ and $V_{imag}=240
  \frac{\hbar^2}{m a_0^2}$ is used for Cs-Cs-Li and Li-Li-Cs. For the
  Cs-Cs-Cs recombination $\rho_{cut}=1.4 a_0$ and $V_{imag}=16
  \frac{\hbar^2}{m a_0^2}$ are used and for Li-Li-Li $\rho_{cut}=0.41
  a_0$ and $V_{imag}=68 \frac{\hbar^2}{m a_0^2}$ are used. }
\label{fig:comparisonofCsLi}
\end{figure}

On figure \ref{fig:comparisonofCsLi} all the recombination
coefficients for the Cs-Li gas have been plotted.  The optical model
parameters are chosen to reproduce the position of the lowest measured
recombination peaks \cite{pires2014,Berninger2011,Dyke2013}.  The
experimental data for Li is actually for $^7$Li \cite{Dyke2013}, but
our aim is here only to test the mass dependency.  In the
$^{133}$Cs-$^6$Li experiment, Li-Li-Li and Li-Li-Cs recombinations are
expected to be suppressed due to Fermi statistics, which is not taken
into account in our model.  So the results are not directly comparable
to the experiment. In addition this is not a direct comparison between
the mass-balanced and mass-imbalanced case, since the recombination
coefficient is a function of different scattering lengths in the
different cases.

Furthermore, for $a_{22}=0$ it was not possible to locate an Efimov
resonance within the range of $a_{12} \in [0,-10000 a_0]$ for neither
the Li-Li-Cs nor the K-K-Rb mass ratio, even though a wide range of
$\rho_{cut}$ were tested.  This is because the Efimov scaling factor
now is so big that it is hard to locate an interval with even one
Efimov resonance.  In these comparisons we shall focus on
recombination coefficients for zero energy.

The absolute sizes on fig.~\ref{fig:comparisonofCsLi} show that the
Cs-Cs-Li recombination is much more likely than the Li-Li-Cs
recombination for the same value of $a_{12}$.  The Cs-Cs-Cs
recombination only depends on $a_{11}$, which is used as the
$x$-coordinate for this process in fig.~\ref{fig:comparisonofCsLi}.
This scattering length is expected to be of less importance compared
to $a_{12}$ in the mixed recombination coefficients, and therefore assumed to
be zero in those estimates.

The comparison is then not straightforward but still useful, since a
finite value of $a_{11}$ would increase $\alpha_{112}^{(1)}$ beyond
the curve in fig.~\ref{fig:comparisonofCsLi}.  Thus we deduce that
$\alpha_{112}^{(1)}(a_{11}=0,a_{12}) \gg
\alpha_{111}^{(1)}(a_{11}=a_{12})$, and we therefore believe that the
Cs-Cs-Li recombination is much more likely to occur than the Cs-Cs-Cs
recombination in realistic systems.  We also see in
fig.~\ref{fig:comparisonofCsLi} that
$\alpha_{222}^{(2)}(a_{22}=a_{12}) \gg
\alpha_{112}^{(2)}(a_{11}=0,a_{12})$.  This does not allow any
conjecture about relative sizes in a realistic system because the
intra-species scattering lengths usually are much smaller than the
necessary large (for the Efimov effect) inter-species scattering
length.  The recombination coefficients between identical particles are then
expected to be relatively small.

\begin{figure}[h]
\centering
\includegraphics[scale=0.25]{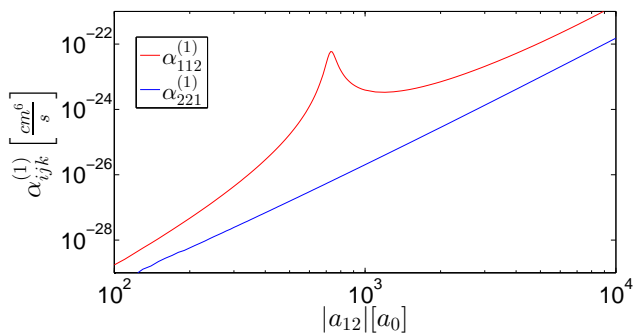}
\caption{Recombination coefficients for $R=2.23$ as a function
  of $a_{12}$. The optical parameters are $\rho_{cut}=2.79 a_0$
  and $V_{imag}=20\frac{\hbar^2}{m a_0^2}$ for both systems. }
\label{fig:comparisonofsystems}
\end{figure}

To complement we now investigate the much
smaller mass ratio $R=2.23$ (corresponding to a Rb-K gas).  The results are 
shown on fig.~\ref{fig:comparisonofsystems} for different recombination
processes.  The optical model parameters are chosen to give a peak for
$a_{12} \approx -700 a_0$ in the $\alpha_{112}^{(1)}$ coefficient.  As for the
larger mass ratio we again conclude that the heavy-heavy-light recombination
process is more likely than the ligt-light-heavy recombination process.

Figs. \ref{fig:comparisonofCsLi} and
\ref{fig:comparisonofsystems} also lead to another conclusion.
A bigger mass-ratio seems to give a bigger recombination coefficient
for two heavy particles and one light.  This is in accordance with a
corresponding decrease of size of the adiabatic potentials, which
intuitively suggests higher probability for recombination at small
hyper-radii.

A bigger mass-ratio also seems to give a somewhat smaller
recombination coefficient for two light particles and one heavy particle,
although not as pronounced.  This means that the bigger the
mass-ratio, the bigger the difference between the heavy-heavy-light
recombination and the light-light-heavy recombination.  For both small
and big mass-ratios, the $\alpha_{221}^{(i)}$ terms can be neglected
in Eqs.(\ref{eq:rateequationalpha1}) and (\ref{eq:rateequationalpha2})
based solely on these mass-related arguments.

We emphasize these conclusions are only strictly valid for small
$|a_{11}|$ and $|a_{22}|$. Finite values imply more complicated
relations between the different recombination coefficients which only can be
determined by taking the Feshbach resonances of a specific system into
account. So in theory, it is possible to have a specific system where
the light-light scattering length $|a_{22}|$ is much bigger than the 
heavy-heavy scattering length $|a_{11}|$, which in turn leads to dominance 
of the light-light-heavy over the heavy-heavy-light process.

\section{Comparison With Experiment}

In this section we will confront our theoretical results with recent
experiments that have been carried out for a $^{133}$Cs-$^{6}$Li gas
\cite{pires2014,tung2014}, as well as with a recent experiment for a
$^{133}$Cs gas, in which a second Efimov resonance has been observed
\cite{Huang2014}. The variable parameter in experiments is the magnetic
field $B$, which via the mechanism of Feshbach resonances can be used
to change the scattering lengths, $a$. The phenomenological relation
between $a$ and $B$ is
\begin{equation}
\label{eq:feshbach}
a(B)=a_{bg}\left(1+\frac{\Delta}{B-B_0}\right)
\end{equation}
where $\Delta$ , $a_{bg}$ and $B_0$ are determined experimentally for
each individual system.  Eq.(\ref{eq:feshbach}) applies for most
systems and will be used in the cases investigated in this section.
We first focus on the equal mass process after which we continue with
the dominating heavy-heavy-light recombination process.

\subsection{The $^{133}$Cs-$^{133}$Cs-$^{133}$Cs recombination}

Recently a second Efimov resonance was observed in a $^{133}$Cs gas
\cite{Huang2014}, and we can conveniently test the model against this
experimental confirmation of the original Efimov scenario.  The first
Efimov resonance for a $^{133}$Cs gas was observed earlier
\cite{Berninger2011} to have a peak for $a^{(-)} \approx -960 a_0$.
The experimental data from these measurements also allow us to adjust
both $\rho_{cut}$ to give the peak position and subsequently tune the
strength, $V_{imag}$, to the shape of this first peak.  With the
scaling mass, $m$, of $^{133}$Cs, $\rho_{cut}=1.4 a_0$ and
$V_{imag}=16 \frac{\hbar^2}{m a_0^2}$, the first Efimov resonance is
then rather well reproduced.

\begin{figure}[h]
\centering
\includegraphics[scale=0.25]{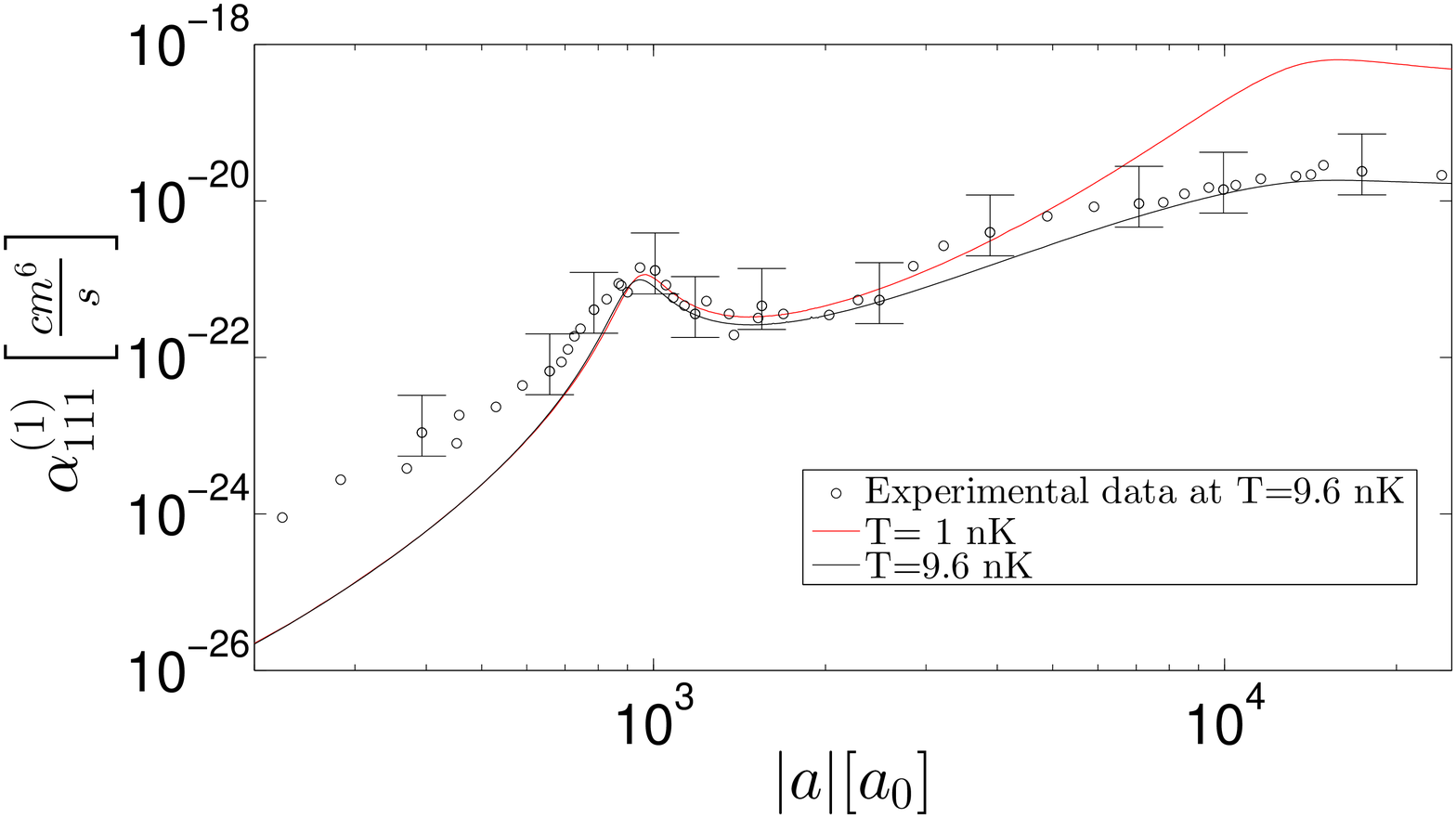}
\caption{The recombination coefficient for Cs-Cs-Cs, $\alpha_{111}^{(1)}$, 
  at different temperatures plotted with experimental data from
  \cite{Berninger2011,Huang2014}.  The optical model parameters are
  $\rho_{cut}=1.4 a_0$ and $V_{imag}=16 \frac{\hbar^2}{m a_0^2}$.  }
\label{fig:absolutescaleCs}
\end{figure} 

We show the calculated results on fig.~\ref{fig:absolutescaleCs},
where the second Efimov peak is obtained without any adjustments
beyond the first peak.  The experimental results are not completely
commensurable, since the first experiment was done at a temperature of
about 15 nK, and the second experiment at 9.6 nK.  We can circumvent
this by using the fact that temperatures of this value are
insignificant at small values of the scattering length.  The data for
the first peak from the first experiment \cite{Berninger2011} can
therefore be assumed to arise for the same temperature of 9.6 nK as
the second peak in the second experiment \cite{Huang2014}. Both
temperatures are much smaller than the upper temperature limit
estimated to be $1 \mu K$ for a realistic calculation.

The absolute value around the first peak and for big $a$ is generally
remarkably close to the experimental value. However, the calculated
recombination coefficient is much too small at small $a$, where the
temperature has no influence.  The theoretical results follow the
overall $a^4$ rule of recombination dependence and these small $a$
deviations must therefore arise from other processes contributing to
the experimental values.

The calculated temperature dependence for larger scattering lengths is
in very good agreement with the measurements.  This is remarkable,
since a reduction in temperature to 1 nK results in a fairly dramatic
change of the recombination curve at large values of $a$.  The
consequence is that a moderate temperature of a few nK already smears
out the second Efimov peak, and prohibits observation.  We then
conclude that using the correct temperature gives a shape that is in
pretty good agreement with the experimental results.  However, the
second peak is dislocated compared to the ideal Efimov scaling factor
of 22.7, which predicts this peak to be at around $a_{12} = -21800~a_0$.  In the
experiment, it is found at approximately $-17000~a_0$, while the
calculation gives the peak at $-15800~a_0$.

Overall, the results of this comparison with experiments for identical
bosons are encouraging. The temperature effects are well accounted for
and the shape and location of the second Efimov peak is also in broad
agreement with data, for phenomenological parameters fitted to the
first peak. Our model seems to work for the well-known
mass-balanced case, and comparison to data for mass-imbalanced systems
should then be considered.

\subsection{The $^{133}$Cs-$^{133}$Cs-$^7$Li recombination}

The crucial parameter is the scattering length, $a_{12} = a_{\mathrm LiCs}$,
which has to be very large to provide the Efimov effect.  However, in
addition also $a_{11} = a_{\mathrm CsCs}$ is important for quantitative
predictions.  The overall $a_{12}^4$ scaling is modified for a finite
value of $a_{11}$, which is determined by the magnetic field through
the Feshbach resonance of the system as described in
Eq.(\ref{eq:feshbach}).  It is then interesting to look for effects in
the recently obtained two sets of experimental data
\cite{pires2014,tung2014}.  They are in broad agreement, although the
details are a little different.  In \cite{tung2014}
3~Efimov resonances are reported, where the third one is very hard to distinguish
from the background due to finite temperature effects.  In
\cite{pires2014} the recombination coefficient is given as a function
of $a_{12}$, which allows an easy comparison with our calculations.

The experimental conditions in \cite{pires2014} provide the parameters
in  Eq.(\ref{eq:feshbach1}) for the interspecies (Cs-Li)
Feshbach-resonance of the prepared spin-states, that is
\begin{equation}
\label{eq:feshbach1}
a_{bg}   = -28.5 a_0 \; ,\; \Delta  = 61.4 ~G  \; ,\; B_0 = 842.9~G \; ,
\end{equation}
which by insertion in Eq.(\ref{eq:feshbach}) gives the variation of
$a_{12}$.  The $a_{11}$ scattering length is estimated to run between
$-1200 a_0 \text{ and } -1500 a_0$ \cite{pires2014}, so we have chosen
a constant intermediate value of $a_{11}=a_{CsCs}=-1350 a_0$.

\begin{figure}[h]
\centering
\includegraphics[scale=0.25]{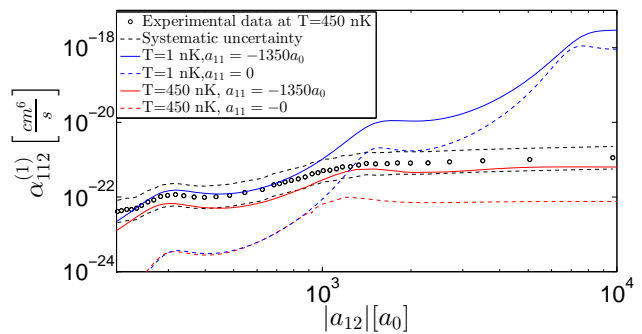}
\caption{The recombination coefficient for Cs-Cs-Li,$\alpha_{112}^{(1)}$
  as a function of the interspecies scattering length, for $a_{11}=0$ and
  $a_{11}=-1350a_0$, plotted with the experimental data from
  \cite{pires2014}.  The optical model parameters are
  $\rho_{cut}=0.258a_0$ , $V_{imag}=100 \frac{\hbar^2}{m a_0^2}$ for
  $a_{11}=-1350 a_0$ and $\rho_{cut}=0.20 a_0$ , $V_{imag}=240
  \frac{\hbar^2}{m a_0^2}$ for $a_{11}=0$.  One small temperature of
  1 nK and the experimental temperature of 450 nK are shown.  }
\label{fig:absolutescale}
\end{figure}

We compare theoretical and experimental results in
fig.~\ref{fig:absolutescale} where both temperature and $a_{11}$
dependence are shown.  We first note that there is a pretty good
correspondence between the experimental data at 450 nK and the
theoretical prediction at 450 nK with the suggested finite value of
$a_{11}$.  The shape is nicely reproduced and the absolute value is
within the systematic uncertainty. We emphasize that the calculations
for vanishing $a_{11}=0$ does not describe the data quantitatively at
450 nK. The absolute value and the shape is only correct when the
enhancement of $\alpha_{112}^{(1)}$ at small values of $a_{12}$, due
to finite values of $a_{11}$, is taken into account. We also draw
attention to the fact that for a substantial finite value of
$|a_{11}|$, $\alpha_{112}^{(1)}$ is no longer independent of energy at
small values of $|a_{12}|$.  The temperature of 450 nK is not far from
$1 \mu K$, that is upper temperature limit in a realistic
calculation. But the data is described very well.

It seems like the calculated values of $\alpha_{ijk}^{(i)}$ describe
the relative behaviour of different recombinations well.  The
recombination coefficient for Cs-Cs-Li is larger than for Cs-Cs-Cs for
both the experimental and the calculated values. The calculated values
of both are within experimental uncertainty.  The relative location of
the Efimov resonances depends on the finite value of $a_{11}$. For the
Cs-Cs-Li system the limiting values are 4.8766 or 4.7989, however,
which means that we do not expect any noteworthy difference. The 3
Efimov resonances in \cite{tung2014} are well suited for testing the
theoretical prediction for the Efimov resonances.

We use the value, $B_0=942.75$, of the resonant magnetic field.
The two reported different values of $B_0$ are within the experimental
uncertainties of both experiments.  In table
\ref{tab:experimentalcomparison}, we compare predicted and
experimental values for the Efimov peaks when the optical model
parameters are adjusted to reproduce either first ($\rho_{cut}=0.258 a_0$) or
second peak ($\rho_{cut}=0.28 a_0$).

\begin{table}[h]
 \caption{The values of $\Delta B =B-B_0$ determined to give the first
   or second Efimov peak for Cs-Cs-Li, while the other two peaks (
   second or first as well as third) are predicted with that set of
   corresponding optical model parameters.}  \tabcolsep=0.11cm
    \begin{tabular}{|l|l|l|l|}
    \hline
                    &   1st peak & 2nd peak     & 3rd peak       \\ \hline
  Experimental value  of $\Delta B$ [G] & 5.61(16) & 1.07(2)  & 0.22(4) \\ \hline
  $\Delta B$ [G] fitted to 2nd peak & 5.8  & 1.07 & 0.21               \\ \hline
  $\Delta B$ [G] fitted to 1st peak & 5.65 & 1.03 & 0.2     \\ \hline
  \end{tabular}
  \label{tab:experimentalcomparison}
\end{table}

Then in both cases the third peak is in agreement with calculations
within the experimental uncertainty.  However, this does not say much,
since a small change in magnetic field results in a huge change in the
scattering length, see Eq.(\ref{eq:feshbach1}) and
Eq.(\ref{eq:feshbach}).  Due to this, it is also hard to determine an
accurate $\rho_{cut}$ value which reproduces the very weak third peak.  The two
predicted peaks in the two fits are at the edge allowed by the
experimental uncertainty, and as such they are still very close to the
observed values. The finite temperature effects are also likely to
move the locations slightly, which might explain smaller deviation.

We can now calculate the inter-species scattering lengths
corresponding to the experimental peak values by using 
Eq.(\ref{eq:feshbach1}) with $B_0=942.75$, which yields
$|a^{(-)}_1|=331.5 a_0$, $|a^{(-)}_2|=1621 a_0$ and $|a^{(-)}_3|=7777.5
a_0$.  This gives a Efimov scaling of 4.8911 between the first two peaks and
4.7980 between the last two peaks.  These values are in agreement with
the Efimov scaling found by going to the universal limits of infinite
scattering lengths, 4.8766 or 4.7989, depending on whether there is
two or three contributing resonant interactions.

\subsection{Universal properties of optical parameters}

For identical bosons ref.\cite{Peder2013} proposes a possible relation
between the optical model parameters and the van der Waals length of
the system. This is motivated by the recent findings in 
equal mass systems \cite{Berninger2011} of a universal relation 
between the three-body parameter and the two-body van der Waals 
length \cite{naidon2011,chin2011,wang2012a,schmidt2012,peder2012,naidon2012,wang2012b}.
We can expand this idea to the mass-imbalanced system. The
three van der Waals lengths which we will compare with
is for Cs-Cs, Li-Li and Cs-Li respectively $202 a_0$, $65 a_0$ and $44.8 a_0$
\cite{Pethich2002,Dervianko2001}.
The crucial parameter is $\rho_{cut}$ which by definition is a
hyper-radius.  From the definition in Eq.(\ref{eq:hypersphericalcoord}) we can
analogously define a hyper-radial van der Waals length as
\begin{equation}
\rho_{vdW}^2=\frac{2R}{1+2 R} \frac{\mu_{11}}{m} r_{vdW,11}^2 +
\frac{1+ R}{1/2 + R}\frac{\mu_{12}}{m}  r_{vdW,12}^2 \; ,
\label{eq:rholength}
\end{equation}
where the distance between particles is replaced by the corresponding
two-body van der Waals lengths, and the related two-body reduced masses,
$\mu_{11}$ and $\mu_{12}$, are introduced and $R=\frac{m_1}{m_2}$.

For identical particles with $m$ equal to the mass of the particles,
this hyper-radius reduces to the van der Waals length, $\rho_{vdW}^2=
r_{vdW,11}^2$, which in ref.\cite{Peder2013} was compared to
$\rho_{cut}$ by simple division. For three identical particles we find
respectively $\rho_{cut}/\rho_{vdW}=0.0069$ and
$\rho_{cut}/\rho_{vdW}=0.0063$ with the parameters for Cs-Cs-Cs and
Li-Li-Li.

For the mass-imbalanced system the (identical) interactions between
the heavy and the two light particles is as dicussed expected to be
the dominating contribution. This means that the last term in
Eq.(\ref{eq:rholength}) should be the largest while the first term
should vanish when the corresponding scattering length, $a_{11}$,
between the identical heavy particles is relatively small.  Furthermore,
when $a_{11}$ is finite and begins to contribute the recombination
peak moves.  To keep the peak location independent of $a_{11}$ we
therefore varied $\rho_{cut}$ slightly with $a_{11}$.  This addition
to $\rho_{cut}$ has to come from the first term of
Eq.(\ref{eq:rholength}) which should vary from zero to a given finite
contribution when $|a_{11}| = \infty$.  Thus we can parametrize by
\begin{eqnarray} 
\rho_{vdW}^2 &=&
\frac{2R}{1+2R} \frac{\mu_{11}}{m} \frac{|a_{11}|}{(f_{11} a_0 + |a_{11}|R^2f(R))}
r_{vdW,11}^2 \nonumber  \\  \label{eq:rholength1}   &+&  
 \frac{1+ R}{1/2+ R}\frac{\mu_{12}}{m}  r_{vdW,12}^2 \; ,
\end{eqnarray}
where we suggest to use $f_{11} = 10^{7}$. For equal masses and $a_{11}\rightarrow\infty$  
this reduces to Eq.\ref{eq:rholength} for equal masses. For unequal masses the $R^2$ factor
ensures that for $a_{11}\rightarrow\infty$, the term still doesn't become dominant, in correspondence
with the moderate change in $\rho_{cut}$, even as big values of $|a_{11}|$ were used numerically.
We use $f(R)=1$, but perhaps a more complicated function is required. 
For the experimental value, $a_{11}=-1350 a_0$ ($\rho_{cut}=0.258 a_0$), 
we then get $\rho_{cut}/ \rho_{vdW}=0.0057$ close to the experimental values for 
identical particles.

The strength, $V_{imag}$, can also be compared to the natural unit for
a van der Waals interaction, $V_{vdW}=\hbar^2/m\rho_{rdW}^2$.  which
amounts to $V_{imag}/V_{vdW}= 6.53 \cdot 10^5, 2.87\cdot 10^5,
2.03\cdot 10^5$ for Cs-Cs-Cs, Li-Li-Li, and Cs-Cs-Li, respectively.
The value of $\frac{m}{\hbar} V_{imag}\rho_{cut}^2$ is 
constant for a given location and shape in a specific system.  The
numerical values from our calculation are for this quantity 31.3600,
11.4308, 6.6564 for the systems Cs-Cs-Cs, Li-Li-Li and Cs-Cs-Li.

The upshot of these comparisons is that a possible approximate value of
$\rho_{cut}$ for a system can be obtained by $\rho_{cut} \approx 0.006
\rho_{vdW}$.  In addition the value of $V_{imag}$ to within a factor
of two seems obtainable by $V_{imag} \approx 3\cdot 10^5
V_{vdW}$. Finally the value $\frac{m}{\hbar} V_{imag}\rho_{cut}^2$ in different
systems are within the same order of magnitude.

\section{Discussion and Outlook}

We have developed a method for calculating the three-body
recombination coefficient of mass-imbalanced three-body systems at
negative values of the scattering lengths and for small energies.  In
order to do this we employed the hyper-spherical method, zero-range
potentials and the Faddev decomposition to calculate the lowest-order
adiabatic potential.

In addition we formulated and explored the optical model to calculate
recombination processes into deep dimers. This method introduces two phenomenological
optical parameters, strength and range, which respectively determine
location and shape of the Efimov resonances. We then developed two methods
for finding the recombination coefficients of both mass-balanced and
mass-imbalanced systems. By means of the traditional S-matrix method
and by a new method, where the decay rate of bound states in a box
due to the presence of the optical potential is calculated. In general the results 
obtained with the two methods are essentially indistinguishable, although 
differing in numerical inaccuracies.  As they tend to complement each other,
we choose the most convenient method in actual calculations.

The model was tested against the experimental data on recombination coefficients
in Cs-Cs-Cs and Cs-Cs-Li systems as functions of the Feshbach tuned scattering lengths.
In both cases, after fitting the range and the strength
of the imaginary potential to the first peak, the model was able to describe
quantitatively the whole curve including the temperature effects.

The two-parameter fits of strength and hyper-radius of the imaginary
part of the optical potential are very efficient for both position and
shape of the recombination peaks. It is also remarkable that the
absolute values of the experimental recombination coefficients are
reproduced within the experimental uncertainty.

Using the developed methods we have reached a number of conclusions.
The main conclusions are that 
recombination is dominated by the heavy-heavy-light process and mainly 
determined by the heavy-light scattering length. But the heavy-heavy 
scattering length is important for obtaining the correct value of the 
recombination coefficients, as it enhances the recombination probability
and modifies the behaviour as a function of the heavy-light scattering length.
For a large mass ratio the Efimov scaling is determined entirely by the heavy-light
scattering length, but for a smaller mass ratio finite values of the
heavy-heavy scattering length becomes important. The Efimov scaling then
moves continuously between values from two and three resonating
subsystems.  

The focus of this paper has been the case where all scattering lengths
are negative, but it could be just as relevant to investigate two
negative, one positive or two positive, one negative scattering length
cases.  Since a finite value of the heavy-heavy scattering length can
substantially alter the absolute value of a given two-component
recombination, it may be interesting to test the three-body recombination
for the same system at different Feshbach resonances experimentally,
in order to further confirm this prediction.

For small mass-ratios we find a big difference in the Efimov scaling between
peaks as the heavy-heavy scattering length is increased. It may be
fruitful to investigate the intermediate area between two and three
resonant subsystems in more theoretical detail.  In addition a
Feshbach resonance in a two-component gas with small mass-ratio, in
which the heavy-heavy scattering length is big in the same area as the
inter-species scattering length would allow investigation of this
effect.

The authors thank PK S{\o}rensen for invaluable help
on the numerical details.
The authors are grateful for enlightening discussions with A.~G. Volosniev,
N. Winter, N.~B. J{\o}rgensen, L.~J. Wacker, J.~F. Sherson and J.~J. Arlt. 
This reserach was supported by the 
Danish Council for Independent Research DFF Natural Sciences.

\end{document}